\documentclass[a4paper,fleqn,usenatbib]{mnras}

\usepackage[T1]{fontenc}
\usepackage{ae,aecompl}

\usepackage{graphicx}	
\usepackage{amsmath}	
\usepackage{amssymb}	
\usepackage{txfonts}

\def\rpd{\hbox{rad\,d$^{-1}$}}

\def\chisq{\hbox{$\chi^2$}}
\def\delchisq{\hbox{$\Delta \chi^2$}}
\def\chisqr{\hbox{$\chi^2_{\rm r}$}}
\def\msun{\hbox{${\rm M}_{\odot}$}}
\def\mjup{\hbox{${\rm M}_{\rm Jup}$}}

\def\rsun{\hbox{${\rm R}_{\odot}$}}

\def\mstar{\hbox{$M_{\star}$}}
\def\rstar{\hbox{$R_{\star}$}}

\def\teff{\hbox{$T_{\rm eff}$}}

\def\sn{\hbox{S/N}}

\def\vraw{\hbox{$v_{\rm raw}$}}
\def\vfilt{\hbox{$v_{\rm fil}$}}
\def\verr{\hbox{$\sigma_{\rm RV}$}}
\def\ms{\hbox{m\,s$^{-1}$}}

\def\kms{\hbox{km\,s$^{-1}$}}
\def\vsini{\hbox{$v \sin i$}}

\def\V{\hbox{${\rm V}$}}
\def\BmV{\hbox{${\rm B-V}$}}
\def\VmRj{\hbox{${\rm V-R_{\rm J}}$}}
\def\VmIj{\hbox{${\rm V-I_{\rm J}}$}}

\def\degr{\hbox{$^\circ$}}

\def\omeq{\hbox{$\Omega_{\rm eq}$}}
\def\dom{\hbox{$d\Omega$}}
\def\Porb{\hbox{$P_{\rm orb}$}}
\def\Prot{\hbox{$P_{\rm rot}$}}
\def\tconj{\hbox{BJD$_{\rm t}$}}

\newcommand{\caii}{\hbox{Ca$\;${\sc ii}}}

\newcommand{\hei}{\hbox{He$\;${\sc i}}}
\newcommand{\hal}{\hbox{H${\alpha}$}}

\title[The hot Jupiter and magnetic activity of V830~Tau]{The hot Jupiter of the magnetically-active weak-line T~Tauri star V830~Tau} 
\author[J.-F.~Donati et al.]{J.-F.~Donati$^{1,2}$\thanks{E-mail: jean-francois.donati@irap.omp.eu},
L.~Yu$^{1,2}$, C.~Moutou$^3$, A.C.~Cameron$^4$, L.~Malo$^{5,3}$, K.~Grankin$^6$, 
\newauthor E.~H\'ebrard$^7$, G.A.J.~Hussain$^{8,1}$, A.A.~Vidotto$^9$, S.H.P.~Alencar$^{10}$, R.D.~Haywood$^{11}$, 
\newauthor J.~Bouvier$^{12,13}$, P.~Petit$^{1,2}$, M.~Takami$^{14}$, G.J.~Herczeg$^{15}$, S.G.~Gregory$^4$, 
\newauthor M.M.~Jardine$^4$, J.~Morin$^{16}$ and the MaTYSSE collaboration 
\newauthor {\it\small  Affiliations are listed at the end of the paper}
}

\date{Submitted 2016 September 16 -- Accepted 2016 November 07} 

\pubyear{2016}

\begin{document}

\label{firstpage}
\pagerange{\pageref{firstpage}--\pageref{lastpage}}
\maketitle

\begin{abstract}
We report results of an extended spectropolarimetric and photometric monitoring of the weak-line T~Tauri star V830~Tau and 
its recently-detected newborn close-in giant planet.  Our observations, carried out within the MaTYSSE programme, were spread 
over 91~d, and involved the ESPaDOnS and Narval spectropolarimeters linked to the 3.6-m Canada-France-Hawaii, 
the 2-m Bernard Lyot and the 8-m Gemini-North Telescopes.
Using Zeeman-Doppler Imaging, we characterize the surface brightness distributions, magnetic topologies and surface 
differential rotation of V830~Tau at the time of our observations, and demonstrate that both distributions evolve with time 
beyond what is expected from differential rotation.  We also report that near the end of our observations, V830~Tau triggered 
one major flare and two weaker precursors, showing up as enhanced red-shifted emission in multiple spectral activity proxies.  

With 3 different filtering techniques, we model the radial velocity (RV) activity jitter (of semi-amplitude 1.2~\kms) that 
V830~Tau generates, successfully retrieve the $68\pm11$~\ms\ RV planet signal hiding behind the jitter, further confirm the 
existence of V830~Tau~b and better characterize its orbital parameters.  
We find that the method based on Gaussian-process regression performs best thanks to its higher ability at modelling not only 
the activity jitter, but also its temporal evolution over the course of our observations, and succeeds at reproducing our RV 
data down to a rms precision of 35~\ms.  
Our result provides new observational constraints on scenarios of star / planet formation and demonstrates the 
scientific potential of large-scale searches for close-in giant planets around T~Tauri stars.  
\end{abstract}




\section{Introduction}
\label{sec:int}

Magnetic fields are thought to play a key role in the formation of stars and their planets \citep[e.g.,][]{Andre09, Baruteau14}, 
and for their subsequent evolution into maturity.  For instance, large-scale fields of low-mass pre-main-sequence (PMS) stars, the 
so-called T~Tauri stars (TTSs), are known to control and even trigger physical processes such as accretion, outflows and angular 
momentum transport, through which they mostly dictate the rotational evolution of TTSs \citep[e.g.,][]{Bouvier07, Frank14}.  
Large-scale fields of TTSs may also help newborn close-in giant planets to avoid falling into their host stars and survive the 
fast migration that accretion discs efficiently trigger, thanks to the magnetospheric gaps that they carve at the disc centre  
\citep[e.g.,][]{Lin96, Romanova06}.  The recent discoveries (or candidate detections) of newborn close-in giant planets around T~Tauri stars 
\citep{vanEyken12, Mann16, JohnsKrull16, Donati16, David16} render the study of the latter topic particularly attractive and timely.  

Although first detected long ago \citep[e.g.,][]{Johns99b, Johns07}, magnetic fields of TTSs are not yet fully characterized, 
neither for those still surrounded by their accretion discs (the classical T~Tauri stars / cTTSs) nor for those whose discs 
have dissipated already (the weak-line T~Tauri stars / wTTSs).  Only recently were the field topologies of a dozen cTTSs unveiled 
\citep[e.g.,][]{Donati07, Hussain09, Donati10b, Donati13} thanks to the MaPP (Magnetic Protostars and Planets) Large Observing 
Programme on the 3.6~m Canada-France-Hawaii Telescope (CFHT) with the ESPaDOnS high-resolution spectropolarimeter (550~hr of clear 
time over semester 2008b to 2012b).  This first exploration revealed for instance that large-scale fields of cTTSs can be either 
relatively simple or quite complex depending on whether the host star is largely convective or mostly radiative \citep{Gregory12, Donati13}; 
it also showed that these fields vary with time \citep[e.g.,][]{Donati11, Donati12, Donati13} and mimic those of mature stars with similar 
internal structures \citep{Morin08b}, suggesting a dynamo origin.  

The ongoing MaTYSSE (Magnetic Topologies of Young Stars and the Survival of close-in giant Exoplanets) Large Programme, allocated at CFHT 
over semesters 2013a-2016b (510~hr) with complementary observations with the Narval spectropolarimeter on the 2-m T\'elescope Bernard Lyot (TBL) 
at Pic du Midi in France (450~hr, allocated) and with the HARPS spectropolarimeter at the 3.6-m ESO Telescope at La Silla in Chile (135~hr, 
allocated), is carrying out the same kind of magnetic exploration on a few tens of wTTSs \citep[][hereafter D14, D15]{Donati14, Donati15}.  MaTYSSE  
also aims at probing the potential presence of newborn close-in giant exoplanets (hot Jupiters / hJs) at an early stage of star / planet formation;  
it recently succeeded at detecting the youngest such body orbiting only 0.057~au (or 6.1 stellar radii) away from the 2~Myr wTTS V830~Tau 
\citep[][hereafter D16]{Donati16}, strongly suggesting that disc migration is a viable and likely efficient mechanism for generating hJs.  

In this new paper, we revisit the latest MaTYSSE data set collected on V830~Tau, including extended observations from early 2016 that follow 
the late 2015 ones from which V830~Tau~b was detected, as well as contemporaneous photometry secured at the Crimean Astrophysical Observatory (CrAO).  
After briefly documenting these additional data (Sec.~\ref{sec:obs}), we apply Zeeman-Doppler Imaging (ZDI) to both subsets to accurately model 
the surface features and large-scale magnetic fields generating the observed activity (Sec.~\ref{sec:mod}).  This modelling is then used to predict 
the activity jitter\footnote{Throughout the paper, we call ``activity jitter'' or ``jitter'' the RV signal that activity generates, and not 
an ``independent, identically distributed Gaussian noise'' as in, e.g., \citet{Aigrain12}.} 
and retrieve the planet signature using two complementary methods, yielding results in agreement with a third completely 
independent technique based on Gaussian-process regression \citep[e.g.,][]{Haywood14, Rajpaul15} and with those of D16 (Sec.~\ref{sec:fil}).  We finally 
summarize our results and stress how MaTYSSE-like explorations can unlock current limitations in our understanding of how giant planets and 
planetary systems form (Sec.~\ref{sec:dis}).

\section{Spectropolarimetric and photometric observations of V830~Tau}
\label{sec:obs}

Following our intensive campaign in late 2015 (D16), V830~Tau was re-observed from 2016 Jan~14 to Feb~10, using again ESPaDOnS at the CFHT, 
its clone Narval at the TBL, and ESPaDOnS coupled to Gemini-North 
through the GRACES fiber link \citep{Chene14}.  ESPaDOnS and Narval collect spectra covering 370 to 1,000~nm at a resolving power of 
65,000 \citep{Donati03}.  A total of 15, 6 and 6 spectra were respectively collected with ESPaDOnS, Narval and ESPaDOnS/GRACES, at a 
daily rate from Jan~14 to 30 and more sparsely afterwards.  ESPaDOnS and NARVAL were used in spectropolarimetric modes, with all collected 
spectra consisting of a sequence of 4 individual subexposures (of duration 690 and 1200~s each for ESPaDOnS and Narval respectively) 
recorded in different polarimeter configurations to allow the removal of all spurious polarisation signatures at first order.  
ESPaDOnS/GRACES spectra were collected in spectroscopic ``star only'' mode, with a resolution similar to that of all other spectra, 
and consist of single 300~s observations.  All raw frames are processed with the reference pipeline {\sc Libre ESpRIT} implementing optimal 
extraction and radial velocity (RV) correction from telluric lines, yielding a typical rms RV precision of 20--30~\ms\ \citep{Moutou07, Donati08b}.  
Least-Squares Deconvolution \citep[LSD,][]{Donati97b} was applied to all spectra, using the same line list as in our previous studies (D15, D16).  
The full journal of observations is presented in Table~\ref{tab:log}.

Rotational and orbital cycles of V830~Tau (denoted $r$ and $o$ in the following equations) are computed from Barycentric Julian Dates (BJDs)
according to the ephemerides:
\begin{eqnarray}
\mbox{BJD} \hbox{\rm ~(d)} & = & 2,457,011.80 + 2.741 r \\
\mbox{BJD} \hbox{\rm ~(d)} & = & 2,457,360.52 + 4.93  o
\label{eq:eph}
\end{eqnarray}
in which the photometrically-determined rotation periods \Prot\ and the orbital period \Porb\ of the hJ are set to 
2.741~d and 4.93~d respectively \citep[][D16]{Grankin13}.  Whereas the initial Julian date of the first ephemeris 
is chosen arbitrarily, that of the second one coincides with the inferior conjunction (with the hJ in front).  

As in our late-2015 data (D16), a few spectra (8 altogether, corresponding to cycles 1.347, 2.090, 2.692, 2.820, 3.068, 3.185, 
3.914 and 4.135) were weakly affected by moonlight in the far blue wing of the spectral lines, due to the proximity of the moon (passing 
within Taurus in Dec and Jan) and / or to non-photometric conditions.  To filter this contamination from our Stokes $I$ LSD profiles, we 
applied the dual-step method described in D16, specifically designed for this purpose and shown to be quite efficient at restoring the 
original RVs down to noise level (50~\ms\ rms in our case, see Table~\ref{tab:log}).  

\begin{table*}
\caption[]{Journal of ESPaDOnS observations of V830~Tau collected in from 2016 Jan~14 to Feb~10.  
ESPaDOnS and Narval spectropolarimetric observations consist of sequences of 4 subexposures (each 
lasting 690~s and 1200~s respectively) whereas ESPaDOnS/GRACES exposures correspond to single (unpolarized) 
observations lasting 300~s each.  Columns $1-5$ respectively list (i)~the UT date of the observation, (ii)~the 
instrument used, (iii)~the corresponding UT time (at mid-exposure), (iv)~the Barycentric Julian Date (BJD), 
and (v)~the peak signal to noise ratio \sn\ (per 2.6~\kms\ velocity bin) of each observation.  
Columns~6 and 7 respectively list the \sn\ in Stokes $I$ LSD profiles (per 1.8~\kms\ velocity bin), and the rms 
noise level (relative to the unpolarized continuum level $I_{\rm c}$) in Stokes $V$ LSD profiles (whenever relevant).  
Columns~8 and 9 indicate the rotational $r$ and orbital $o$ cycles associated with each exposure (using the 
ephemerides given by Eq.~\ref{eq:eph}).  Columns 10--12 respectively give the raw and ZDI-filtered RVs \vraw\ and 
\vfilt, as well as the corresponding 1$\sigma$ error bars \verr.  No \vfilt\ estimates are available for Jan~30 
and Feb~10 spectra, affected by strong flares.  The observation log of our late 2015 data can be found in D16 
(Extended Data Table 1).  }
\begin{tabular}{cccccccccccc}
\hline
Date   & Instrument & UT      & BJD      & \sn\ & $\sn_{\rm LSD}$ & $\sigma_{\rm LSD}$ & $r$ & $o$ & \vraw\ & \vfilt\ & \verr\ \\
(2016) &            & (hh:mm:ss) & (2,457,400+) &      &     &   (0.01\%)  & (142+)      & (8+)        & (km/s) & (km/s)    & (km/s)  \\
\hline
Jan 14 & ESPaDOnS  & 08:19:57 &  1.85135 & 150 & 1460 & 3.3 &  0.303 & 0.384 &   0.254  & $-0.017$ & 0.049 \\
Jan 15 & ESPaDOnS  & 08:16:30 &  2.84889 & 150 & 1400 & 3.3 &  0.667 & 0.586 &   0.789  &   0.020  & 0.051 \\
Jan 16 & ESPaDOnS  & 08:34:49 &  3.86153 & 160 & 1480 & 2.9 &  1.036 & 0.791 & $-0.287$ &   0.005  & 0.048 \\
Jan 17 & ESPaDOnS  & 05:02:34 &  4.71408 & 170 & 1470 & 2.9 &  1.347 & 0.964 & $-0.008$ &   0.000  & 0.049 \\
Jan 18 & ESPaDOnS  & 07:32:40 &  5.81823 & 160 & 1420 & 3.1 &  1.750 & 1.188 & $-0.380$ & $-0.016$ & 0.050 \\
Jan 19 & ESPaDOnS  & 05:55:30 &  6.75069 & 170 & 1470 & 2.9 &  2.090 & 1.377 & $-0.123$ & $-0.092$ & 0.049 \\
Jan 20 & Narval    & 21:30:38 &  8.39998 &  90 & 1130 & 5.1 &  2.692 & 1.712 &   0.546  &   0.089  & 0.063 \\
Jan 21 & ESPaDOnS  & 05:55:40 &  8.75065 & 150 & 1440 & 3.3 &  2.820 & 1.783 & $-1.013$ &   0.034  & 0.050 \\
Jan 21 & Narval    & 22:16:00 &  9.43141 & 100 & 1240 & 4.6 &  3.068 & 1.921 & $-0.099$ &   0.035  & 0.058 \\
Jan 22 & ESPaDOnS  & 05:56:39 &  9.75126 & 140 & 1450 & 3.5 &  3.185 & 1.986 &   0.386  & $-0.006$ & 0.049 \\
Jan 23 & ESPaDOnS  & 07:00:55 & 10.79581 & 160 & 1450 & 3.0 &  3.566 & 2.198 &   1.170  &   0.015  & 0.050 \\
Jan 24 & ESPaDOnS  & 05:57:09 & 11.75144 & 170 & 1450 & 3.0 &  3.914 & 2.392 & $-1.258$ & $-0.069$ & 0.049 \\
Jan 24 & Narval    & 20:25:57 & 12.35475 &  70 &  970 & 6.5 &  4.135 & 2.514 &   0.180  &   0.025  & 0.074 \\
Jan 25 & ESPaDOnS  & 07:23:59 & 12.81166 & 150 & 1470 & 3.4 &  4.301 & 2.607 &   0.284  & $-0.011$ & 0.049 \\
Jan 26 & ESPaDOnS  & 06:59:05 & 13.79429 & 150 & 1420 & 3.5 &  4.660 & 2.806 &   0.840  & $-0.011$ & 0.051 \\
Jan 26 & Narval    & 19:34:05 & 14.31857 &  90 & 1130 & 5.4 &  4.851 & 2.912 & $-1.160$ &   0.039  & 0.064 \\
Jan 27 & ESPaDOnS  & 06:05:23 & 14.75691 & 170 & 1470 & 2.9 &  5.011 & 3.001 & $-0.482$ &   0.009  & 0.049 \\
Jan 28 & ESPaDOnS  & 06:05:42 & 15.75705 & 160 & 1440 & 3.1 &  5.376 & 3.204 & $-0.176$ & $-0.031$ & 0.050 \\
Jan 29 & ESPaDOnS  & 06:58:43 & 16.79378 & 150 & 1410 & 3.3 &  5.754 & 3.415 & $-0.444$ & $-0.036$ & 0.051 \\
Jan 29 & Narval    & 20:01:53 & 17.33762 &  80 & 1210 & 5.6 &  5.952 & 3.525 & $-0.956$ &   0.011  & 0.059 \\
Jan 30 & Narval    & 20:15:57 & 18.34730 &  80 & 1190 & 6.1 &  6.321 & 3.730 &   0.000  &          & 0.060 \\
Feb 04 & GRACES    & 07:12:12 & 22.80262 & 140 & 1560 &     &  7.946 & 4.633 & $-1.017$ & $-0.005$ & 0.046 \\
Feb 04 & GRACES    & 07:18:12 & 22.80678 & 150 & 1560 &     &  7.948 & 4.634 & $-0.978$ &   0.023  & 0.046 \\
Feb 09 & GRACES    & 07:16:45 & 27.80532 & 150 & 1470 &     &  9.771 & 5.648 & $-0.621$ & $-0.019$ & 0.049 \\
Feb 09 & GRACES    & 07:22:39 & 27.80941 & 150 & 1470 &     &  9.773 & 5.649 & $-0.628$ & $-0.010$ & 0.049 \\
Feb 10 & GRACES    & 05:21:11 & 28.72497 & 160 & 1600 &     & 10.107 & 5.835 & $-0.653$ &          & 0.045 \\
Feb 10 & GRACES    & 05:27:05 & 28.72907 & 160 & 1590 &     & 10.108 & 5.836 & $-0.647$ &          & 0.045 \\
\hline
\end{tabular}
\label{tab:log}
\end{table*}

Contemporaneous BVR$_{\rm J}$I$_{\rm J}$ photometric observations were also collected from the CrAO 1.25~m telescope 
(see Table~\ref{tab:pho}), showing that V830~Tau exhibited significantly larger brightness fluctuations than a year 
before (D15), with a full amplitude of 0.28~mag and a period of $2.7424\pm0.0014$~d \citep[compatible within 
error bars with the average periods of][used to phase our spectroscopic data, see Eq.~\ref{eq:eph}]{Grankin13}.

\begin{table}
\caption[]{Journal of contemporaneous CrAO multicolour photometric observations of V830~Tau collected from 2015 Oct~30 to 
2016 Mar~15, respectively listing the Heliocentric Julian Date (HJD) of the observation, the measured \V\ magnitude,
\BmV, \VmRj\ and \VmIj\ Johnson photometric colours, and the corresponding rotational cycle (using again the ephemerides of 
Eq.~\ref{eq:eph}).  The middle line separates observations collected in 2015 and 2016.  The typical 1$\sigma$ error bar 
on \V\ is 20~mmag.  }
\begin{tabular}{cccccc}
\hline
HJD          & \V    & \BmV  & \VmRj & \VmIj & $r$ \\
(2,457,300+) & (mag) & (mag) & (mag) & (mag) & (114+) \\
\hline
 26.4574 & 12.410 &       & 1.339 & 2.182 &  0.797 \\
 28.5085 & 12.378 &       &       & 2.242 &  1.545 \\
 30.6098 & 12.322 & 1.375 & 1.325 & 2.151 &  2.311 \\
 31.5960 & 12.474 & 1.413 & 1.341 & 2.231 &  2.671 \\
 32.5964 & 12.267 & 1.349 &       & 2.143 &  3.036 \\
 40.5358 & 12.307 &       & 1.340 & 2.165 &  5.933 \\
 44.4530 & 12.321 &       & 1.321 & 2.146 &  7.362 \\
 47.5027 & 12.342 &       & 1.307 & 2.162 &  8.475 \\
 47.5524 & 12.367 &       & 1.335 & 2.180 &  8.493 \\
 73.3208 & 12.317 &       & 1.319 & 2.161 & 17.894 \\
 73.5082 & 12.261 &       & 1.305 & 2.130 & 17.962 \\
 74.2657 & 12.268 &       & 1.310 & 2.128 & 18.238 \\
\hline
 91.3173 & 12.281 &       & 1.307 & 2.122 & 24.459 \\
101.2599 & 12.199 &       & 1.282 & 2.094 & 28.087 \\
105.2842 & 12.356 &       & 1.319 & 2.155 & 29.555 \\
112.2820 & 12.244 &       & 1.284 & 2.113 & 32.108 \\
118.2602 & 12.306 &       & 1.339 & 2.160 & 34.289 \\
127.2569 & 12.362 &       & 1.325 & 2.165 & 37.571 \\
129.2079 & 12.252 &       & 1.296 & 2.105 & 38.283 \\
141.2201 & 12.413 &       & 1.343 & 2.204 & 42.665 \\
142.2211 & 12.210 &       & 1.301 & 2.106 & 43.031 \\
153.2096 & 12.219 &       & 1.297 & 2.106 & 47.040 \\
156.2255 & 12.196 &       & 1.288 & 2.094 & 48.140 \\
158.2696 & 12.329 &       & 1.332 & 2.148 & 48.886 \\
163.2378 & 12.425 &       & 1.322 & 2.217 & 50.698 \\
\hline
\end{tabular}
\label{tab:pho}
\end{table}

We note that V830~Tau features emission in various spectral activity proxies, as expected from its youth and fast rotation.  
More specifically, Balmer lines and in particular \hal, are in emission, as well as the central core of the \caii\ 
infrared triplet (IRT) lines, with typical equivalent widths of 85 and 16~\kms\ for \hal\ and the \caii\ IRT emission 
core respectively.  The \hei\ $D_3$ line is most of the time quite shallow, with an average equivalent width of 5~\kms.  

\begin{figure}
\includegraphics[scale=0.35,angle=-90]{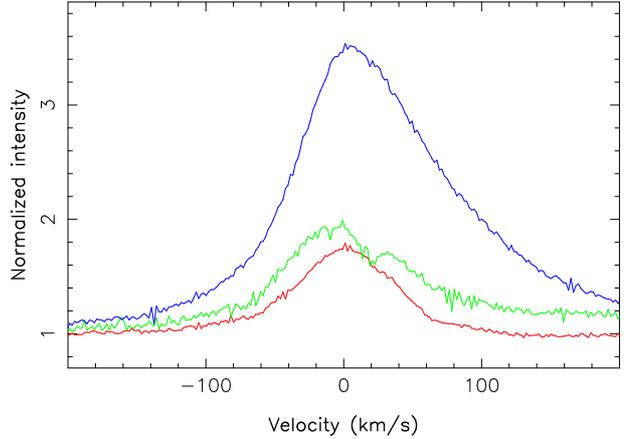} 
\caption[]{\hal\ profiles of V830~Tau on 2016 Jan~17 (cycle 1.347, red line), Feb~10 (cycle 10.107, blue) and Jan.~30 
(cycle 6.321, green).  A red component / tail is clearly present in the latter two profiles (recorded during a flare) 
while absent in the first one (more typical of V830~Tau). } 
\label{fig:hal}
\end{figure}

In 2016 however, we detected several flares of V830~Tau, showing up as enhanced red-shifted emission in all activity 
proxies including \hei, a reliable proxy whose high excitation potential makes it possible to separate flares from phases 
of enhanced chromospheric activity \citep[e.g.,][]{Montes97}.  
The most intense flare occurred on Feb~10 during our last pair of observations (cycles 10.107 and 10.108), when \hal, 
\caii\ IRT and \hei\ emission reach equivalent widths of 280, 32 and 25~\kms\ and feature large red-shifts of 15--35~\kms\ 
(with respect to the stellar rest frame, shifted from the Barycentric rest frame by $\simeq$17~\kms) and asymmetric profiles 
(with a conspicuous red tail for \hal, see Fig.~\ref{fig:hal}, and \hei).  
We note that one of our photometric measurements was secured just after this large flare (at rotation cycle 152.283, or 
10.283 in the reference frame of Table~\ref{tab:log}).  At this time, the star was observed to be 54~mmag (i.e., 2.7$\sigma$) 
brighter than 4 rotation cycles earlier at almost the same phase (cycle 148.289, see Table~\ref{tab:pho}).  This shows 
that even the largest flare of our run was barely detectable in the light curve, to the point that it is not even clear 
which of the two photometric measurements at this phase deviates most from the bulk of our data points (see Sec.~\ref{sec:mod}).  

A weaker flare was detected 10.3~d earlier on Jan~30 (cycle 6.321), with activity proxies exhibiting similar albeit less 
drastic characteristics, e.g., \hei\ emission with an equivalent width of 11~\kms\ and a redshift of $\simeq$20~\kms\ 
(with respect to the stellar rest frame, or $\simeq$10 \kms\ with respect to the average velocity of the \hei\ line).  
A third flare was recorded on Jan~26 (cycle 4.851), mostly in \hal\ (with an equivalent width reaching 122~\kms), but 
short enough to be seen only with Narval, but neither a few hours before (cycle 4.660) nor later (cycle 5.011) with ESPaDOnS;  
this flare has only mild \hei\ characteristics however, with an equivalent width only slightly above average and no 
significant redshift (with respect to the average line velocity).  

The 3 Stokes $I$ spectra corresponding to the 2 first flares turned out to yield discrepant RV estimates (with excess 
blue-shifts of order 0.3~\kms), most likely as 
a result of flaring, and were removed from the subsequent modelling (see Secs.~\ref{sec:mod} and \ref{sec:fil}).  
The Stokes $V$ spectrum associated with the second flare compares well with those collected at similar phases but previous
cycles (0.303, 4.301), suggesting that it was largely unaffected by the flare and thus used for magnetic imaging (see Sec.~\ref{sec:mod}).  
The Stokes $I$ (and $V$) spectra corresponding to the third, milder, flare, yielding an RV estimate consistent with those 
from the two unperturbed ESPaDOnS spectra bracketing the flare, were also kept in the sample.  

\begin{figure*}
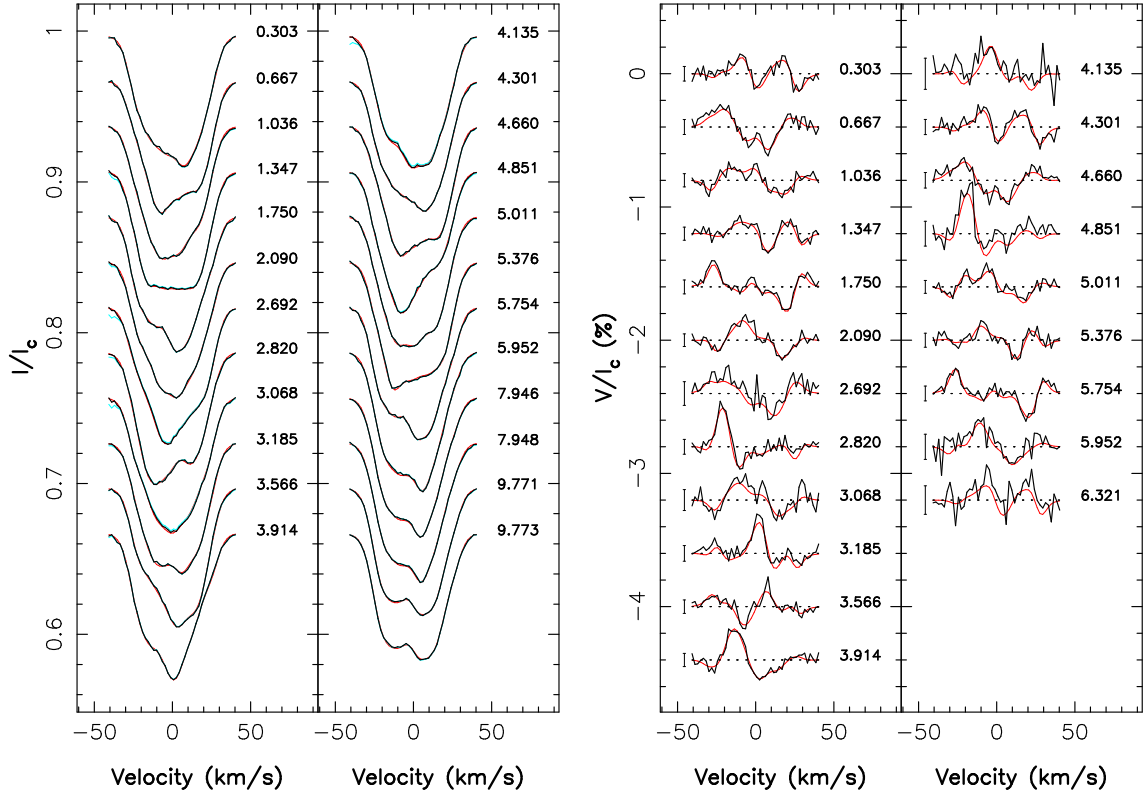

\center{\hbox{\hspace{10mm}
\includegraphics[scale=0.6,angle=-90]{fig/v830_fiti16.ps}\hspace{3mm}
\includegraphics[scale=0.6,angle=-90]{fig/v830_fitv16.ps}}}
\caption[]{Maximum-entropy fit (thin red line) to the observed (thick black line) Stokes $I$ (left panel) 
and Stokes $V$ (right panel) LSD photospheric profiles of V830~Tau in early 2016.  (The red and black lines 
almost perfectly overlap for Stokes $I$ LSD profiles.)  Stokes $I$ LSD profiles 
prior to their filtering from lunar contamination (in the far blue wing) are also shown (cyan line).  
Rotational cycles and 3$\sigma$ error bars (for Stokes $V$ profiles) are also shown next to each profile. }
\label{fig:fit16}
\end{figure*}

\section{Tomographic modelling of surface features, magnetic fields and activity}
\label{sec:mod}

We applied ZDI to both our late-2015 and early-2016 sets of phase-resolved Stokes $I$ and $V$ LSD profiles, keeping 
them separate from each other in a first step.  ZDI is a tomographic technique inspired from medical imaging, with 
which distributions of brightness features and magnetic fields at the surfaces of rotating stars can be reconstructed from 
time-series of high-resolution spectropolarimetric observations \citep{Brown91, Donati97c, Donati01, Donati06b}.  Technically 
speaking, ZDI follows the principles of maximum-entropy image reconstruction, and iteratively looks for the image with lowest 
information content that fits the data at a given \chisq\ level.  By working out the amount of latitudinal shearing that surface 
maps are subject to as a function of time, ZDI can also infer an estimate of differential rotation at photospheric level 
\citep{Donati97a, Donati03b}.  

For this study, we used the latest implementation of ZDI, where the large-scale field is decomposed into its poloidal and 
toroidal components, both expressed as spherical harmonics expansions \citep{Donati06b}, and where the brightness distribution 
incorporates both cool spots and warm plages\footnote{In this paper, the term ``plage'' refers to a photospheric region 
brighter than the quiet photosphere, and not to a bright region at chromospheric level (as in solar physics). } 
(D14, D15, D16).  The local Stokes $I$ and $V$ profiles are computed 
using Unno-Rachkovsky's analytical solution to the polarized radiative transfer equations in a Milne-Eddington model atmosphere, 
taking into account the local brightness and magnetic field;  these local profiles are then integrated over the visible hemisphere 
to derive the synthetic profiles of the rotating star, to be compared with our observations.  This computation scheme provides 
a reliable description of how line profiles are distorted in the presence of magnetic fields \citep[including magneto-optical
effects, e.g.,][]{Landi04}.

In this new paper, we assume for V830~Tau the same parameters as in our previous studies in particular an inclination of the rotation 
axis to the line of sight $i$ equal to $55\pm10$\degr\ and a line-of-sight-projected equatorial rotation velocity \vsini\ equal to 
$30.5\pm0.5$~\kms\ (D15, D16)\footnote{The distance assumed for V830~Tau in D15 and D16, i.e., $131\pm3$~pc, is likely underestimated, 
since V830~Tau is located in L1529 rather than L1495, and thus close to DG~Tau for which the adopted distance is $150\pm5$~pc \citep{Rodriguez12}.  
Given that this difference in distance is comparable flux-wise to the uncertainty on the unspotted magnitude of V830~Tau, we still assume 
for V830~Tau the same stellar parameters as in D15 and D16 (see Table~\ref{tab:spar}).  Assuming instead that V830~Tau is 30\% brighter 
would mostly imply that it is younger, with an age of $\simeq$1.5~Myr (using the evolutionary models of \citealt{Siess00} as in D15).}.  
We recall that the inclination angle $i$ is derived both from the measured stellar parameters (see Sec.~3 of D15) and by minimizing 
the information content of reconstructed images, with a typical error bar of order 10\degr.  
We further assume that the (weak) surface differential rotation of V830~Tau is as derived by D16 from our late 2015 data alone, before 
revisiting the subject using the whole data set in Sec.~\ref{sec:dr}.  The parameters of V830~Tau used in our study are summarized 
in Table~\ref{tab:spar}.  

\begin{table}
\caption[]{Summary of the main parameters of V830~Tau, with references as mentioned whenever appropriate (G13 and R12 stand for 
\citealt{Grankin13} and \citealt{Rodriguez12}).}
\center{
\begin{tabular}{ccc}
\hline
Parameter            & Value            & Reference \\ 
\hline
\mstar\ (\msun)      & $1.00\pm0.05$       & D15 \\ 
\rstar\ (\rsun)      & $2.0\pm0.2$         & D15 \\ 
age (Myr)            & $\simeq$2.2         & D15 \\ 
\Prot\ (d)           & 2.741               & G13 \\ 
BJD$_0$              & 2,457,011.80        & D15 \\ 
\omeq\ (\rpd)        & $2.29525\pm0.00020$ & D16 \\
\dom\ (\rpd)         & $0.0172\pm0.0014$   & D16 \\
$i$ (\degr)          & $55\pm10$           & D15 \\ 
\vsini\ (\kms)       & $30.5\pm0.5$        & D15 \\ 
distance (pc)        & $150\pm5$           & R12 \\ 
\teff (K)            & $4250\pm50$         & D15 \\ 
\hline
\end{tabular}} 
\label{tab:spar}
\end{table}

\subsection{Brightness and magnetic imaging}

\begin{figure*}
\hbox{\hspace{15mm}
\includegraphics[scale=0.45,angle=-90]{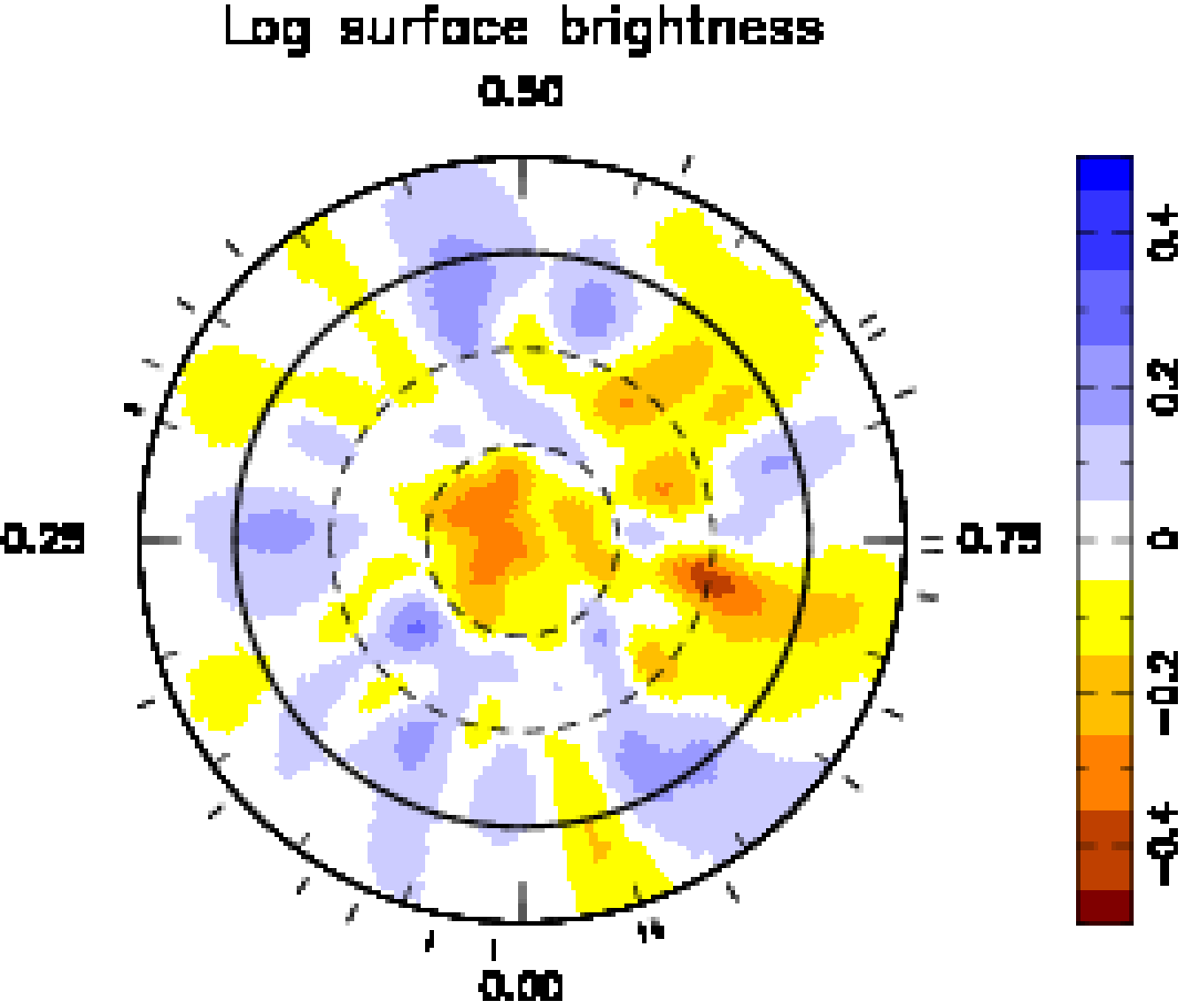}\hspace{20mm}
\includegraphics[scale=0.45,angle=-90]{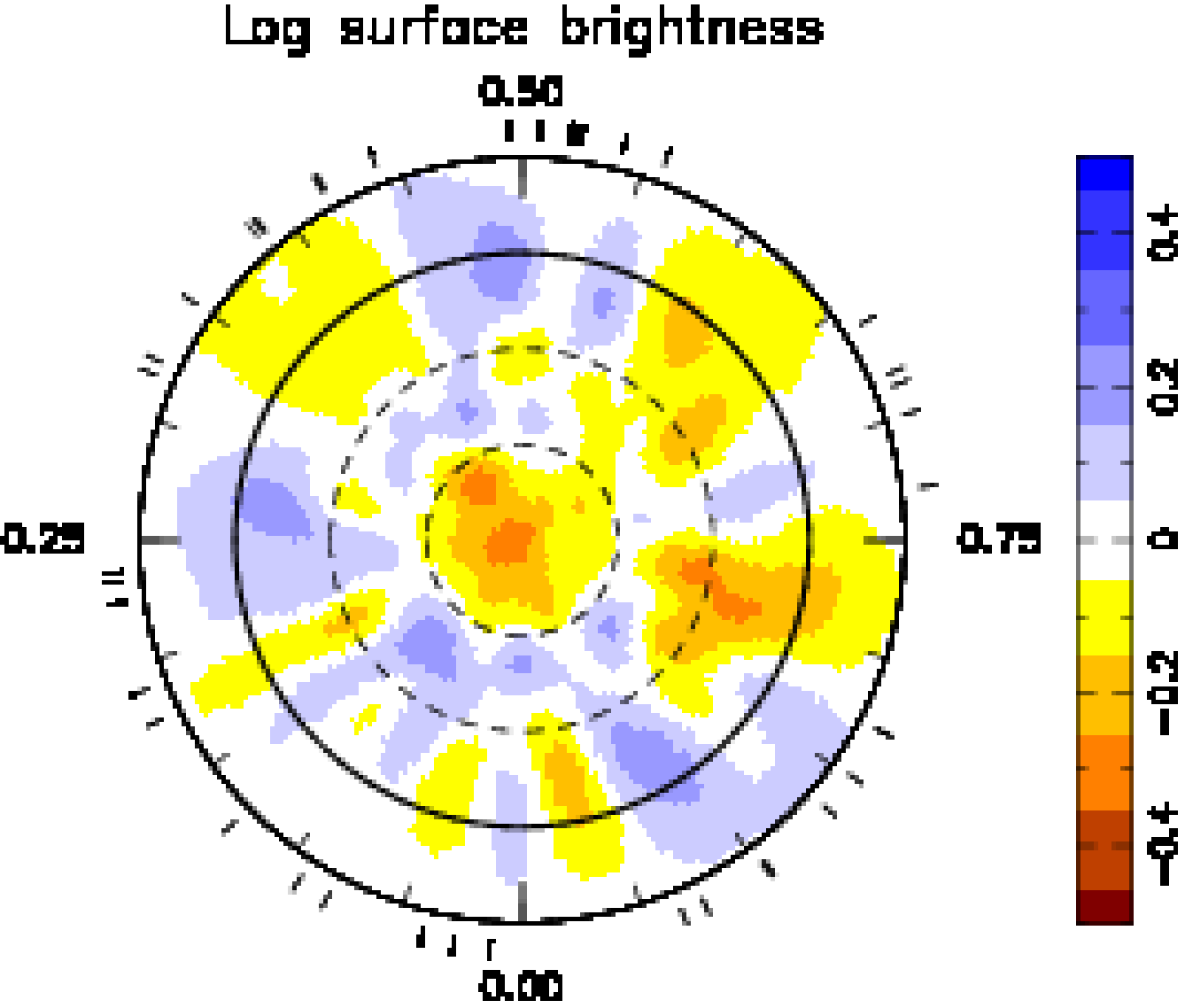}} 
\caption[]{Maps of the logarithmic brightness (relative to the quiet photosphere), at the surface of V830~Tau in 
early 2016 (left) and late 2015 (right).  Cool spots / bright plages show up as brown / blue features. 
The star is shown in flattened polar projection down to latitudes of $-30\degr$,
with the equator depicted as a bold circle and parallels as dashed circles.  Radial ticks around
each plot indicate phases of observations. }
\label{fig:mapi}
\end{figure*}

\begin{figure*}
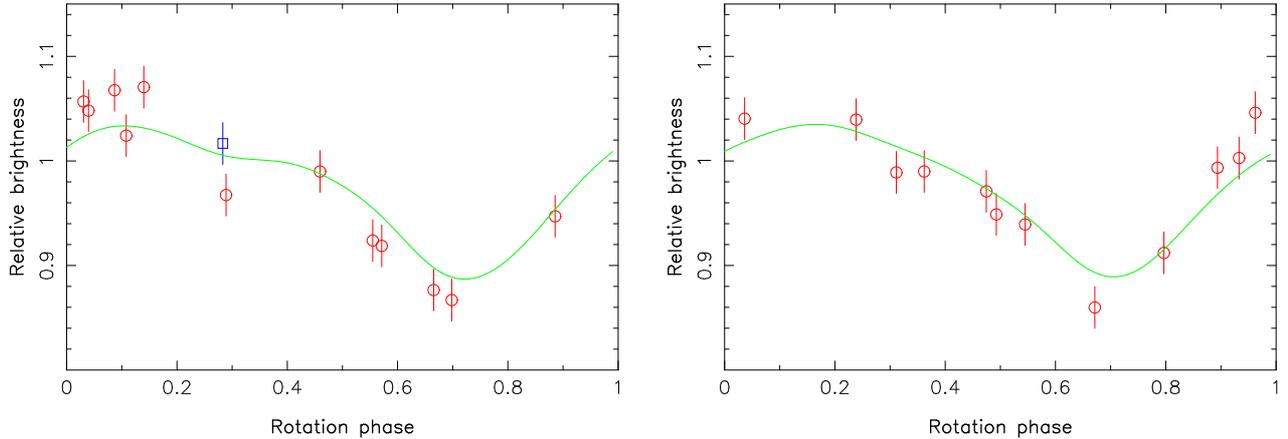

\hbox{\hspace{4mm}\includegraphics[scale=0.35,angle=-90]{fig/v830_pho16_2.ps}\hspace{6mm}\includegraphics[scale=0.35,angle=-90]{fig/v830_pho15.ps}} 
\caption[]{Brightness variations of V830~Tau in early 2016 (left) and late 2015 (right) as predicted from the tomographic modelling 
of our spectropolarimetric data (see Fig.~\ref{fig:mapi}, green line), compared with contemporaneous photometric observations 
in the \V\ band (open symbols and 1$\sigma$ error bars of 20~mmag) at the 1.25-m CrAO telescope (see Table~\ref{tab:pho}).  
The photometric measurement collected immediately after the large flare detected in our early 2016 spectroscopy data (see Sec.~\ref{sec:obs}) 
is shown as an open blue square in the left panel.   }
\label{fig:pho}
\end{figure*}

In Fig.~\ref{fig:fit16}, we show our sets of Stokes $I$ and $V$ LSD profiles of V830~Tau from early 2016, along with the fit to the 
data.  A similar plot is provided in Appendix~\ref{sec:fig} for our late-2015 data set (see Fig.~\ref{fig:fit15}, repeating Fig.~1 
of D16 for Stokes $I$ profiles, and including Stokes $V$ profiles not previously shown in D16).  
The fit we obtain in both cases corresponds to a \chisq\ equal to the number of data points, i.e., to a unit \chisqr\ 
level (where \chisqr\ is simply taken here as \chisq\ divided by the number of data points\footnote{This is the usual convention in 
regularized tomographic imaging techniques where the number of model parameters, reflecting the (ill-defined) number of resolution 
elements in the reconstructed image, is much smaller than the number of fitted data points and not taken into account in the expression 
of \chisqr. }, respectively equal to 1104 and 2208 for 
the early-2016 and late-2015 Stokes $I$ data sets, and to 966 and 1472 for the corresponding Stokes $V$ data sets).  
The initial \chisqr\ values, corresponding to input maps with null fields and no brightness features, are equal to 27 and 19 for the 
early-2016 and late-2015 data sets respectively, clearly demonstrating the overall success of ZDI at modelling the observed modulation of 
both Stokes $I$ and $V$ LSD profiles. 

The reconstructed brightness maps of V830~Tau at both epochs are shown in Fig.~\ref{fig:mapi}.  The two maps share obvious similarities 
and exhibit similar spottedness levels, i.e., $\simeq$13\%\ of the stellar 
surface\footnote{We stress that ZDI is only sensitive to large brightness features, and not to small ones evenly distributed at 
the surface of the star;  for this reason, the value we quote here for the spot coverage of V830~Tau is likely to be a lower limit, 
in agreement with photometric monitoring suggesting a typical spot coverage in the range 30--50\% for V830~Tau \citep{Grankin08}.}  
(7\%\ and 6\%\ for cool and warm features respectively). 
In particular, most cool spots and warm plages present in either maps are recovered at both epochs.  
One can also notice differential rotation shearing the brightness distribution between late 2015 and early 2016 (a time gap corresponding to 
49~d or 18 rotation cycles), with equatorial and polar features being both shifted by a few \% of a rotation cycle to smaller and larger phases 
respectively\footnote{For instance, the equatorial plage at phase 0.46 in early 2016 is found at phase 0.48 in late 2015, while the cool polar 
cap rotated by $\simeq$0.1~cycle in the other direction at the same time.  Note that the latest map is shown first in Fig.~\ref{fig:mapi} and 
following plots.} (implying a fast equator and a slow pole, in good quantitative agreement 
with D16).  Some intrinsic temporal evolution beyond differential rotation may be visible in our images as well, with, e.g., the appearance of a 
warm equatorial plage at phase 0.08 in early 2016 that was not visible (or not as strong) in late 2015;  however, even though phase 
coverage is fairly good in our case at both epochs, quantifying spot evolution by visually comparing images derived from differently 
sampled data sets is notoriously ambiguous and misleading.  We come back on this point in Sec.~\ref{sec:dr}. 

We stress that the derived brightness images predict light curves that are in good agreement with our observations (see 
Fig.~\ref{fig:pho}), even though these images were produced from our sets of LSD profiles only.  Note the small temporal evolution 
in the predicted light curves between both epochs, that our photometric observations cannot confirm due to their limited sampling 
and precision.  This further demonstrates that LSD profiles contain enough information to accurately predict the surface distribution 
of brightness features, and in particular those responsible of the RV activity jitter (see Sec.~\ref{sec:fil});  on the opposite, it 
is quite obvious that photometric information is way too limited (even when better sampled and more precise) to infer complex spot 
distributions such as those we reconstruct for V830~Tau.  It implies that jitter-filtering techniques based solely on photometry 
\citep[e.g.,][]{Aigrain12} are likely to yield poorer results, especially for moderate to fast rotators whose optical RV curves are 
much more sensitive than photometry to small features in surface brightness distributions.  

\begin{figure*}
\vspace{0mm}
\hbox{\hspace{5mm}\includegraphics[scale=0.65,angle=-90]{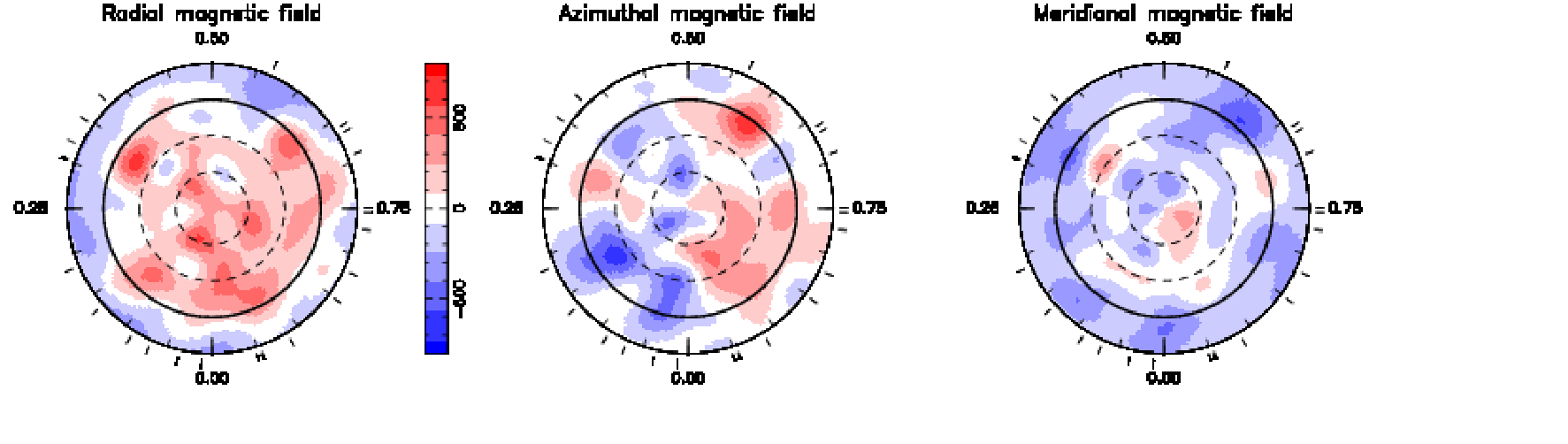}\vspace{-2mm}} 
\hbox{\hspace{5mm}\includegraphics[scale=0.65,angle=-90]{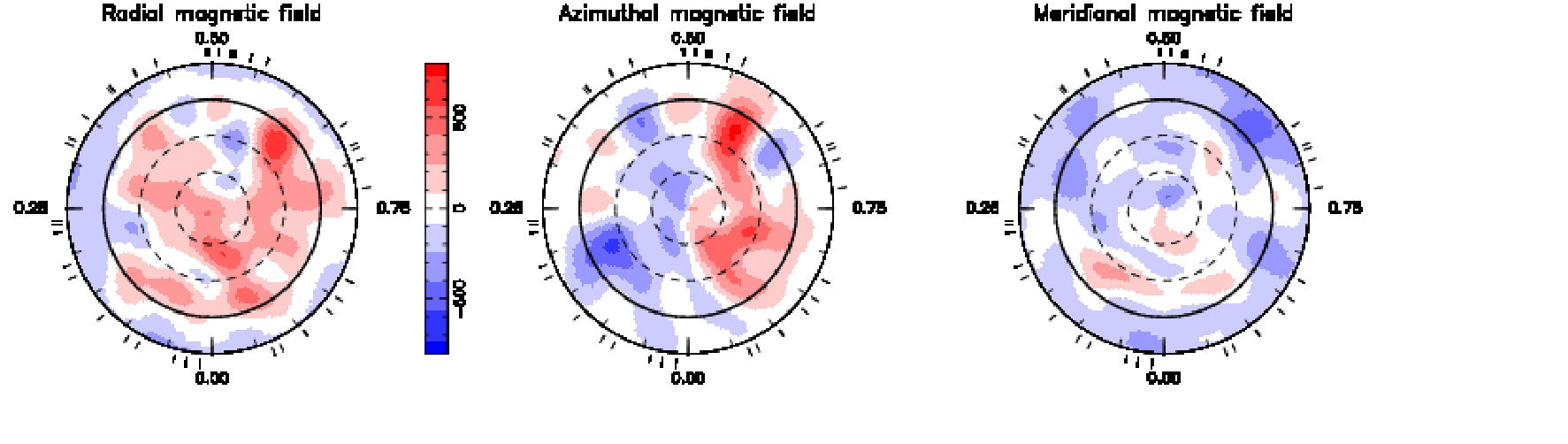}\vspace{-2mm}}
\vspace{-2mm}
\caption[]{Maps of the radial (left), azimuthal (middle) and meridional (right) components of the 
magnetic field $\bf B$ at the surface of V830~Tau in early 2016 (top) and late 2015 (bottom).  
Magnetic fluxes in the color lookup table are expressed in G.  The star is shown in flattened polar 
projection as in Fig.~\ref{fig:mapi}. }
\label{fig:mapv}
\end{figure*}

\begin{figure*}
\hbox{\includegraphics[scale=0.55,angle=-90]{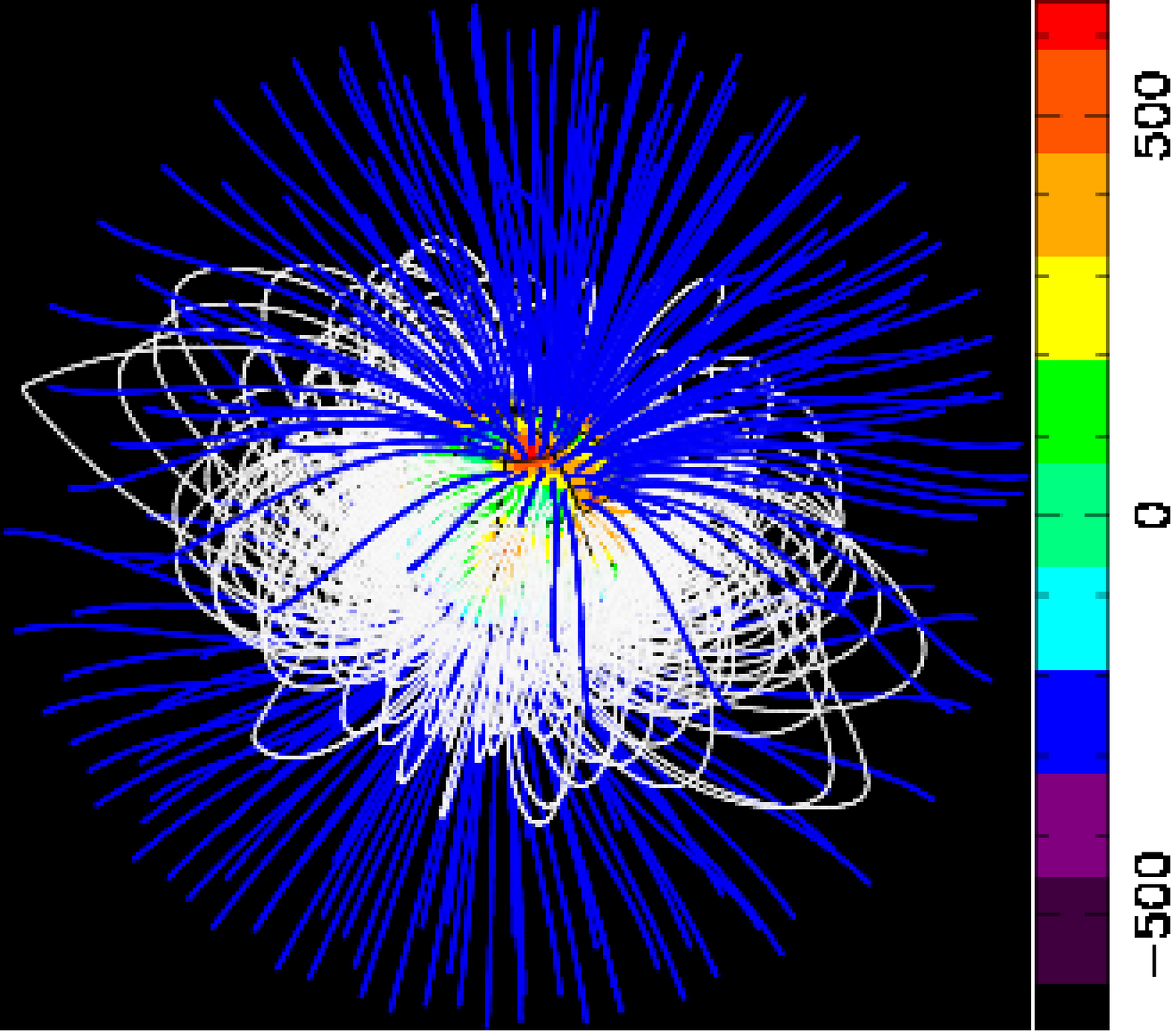}\hspace{3mm}
      \includegraphics[scale=0.55,angle=-90]{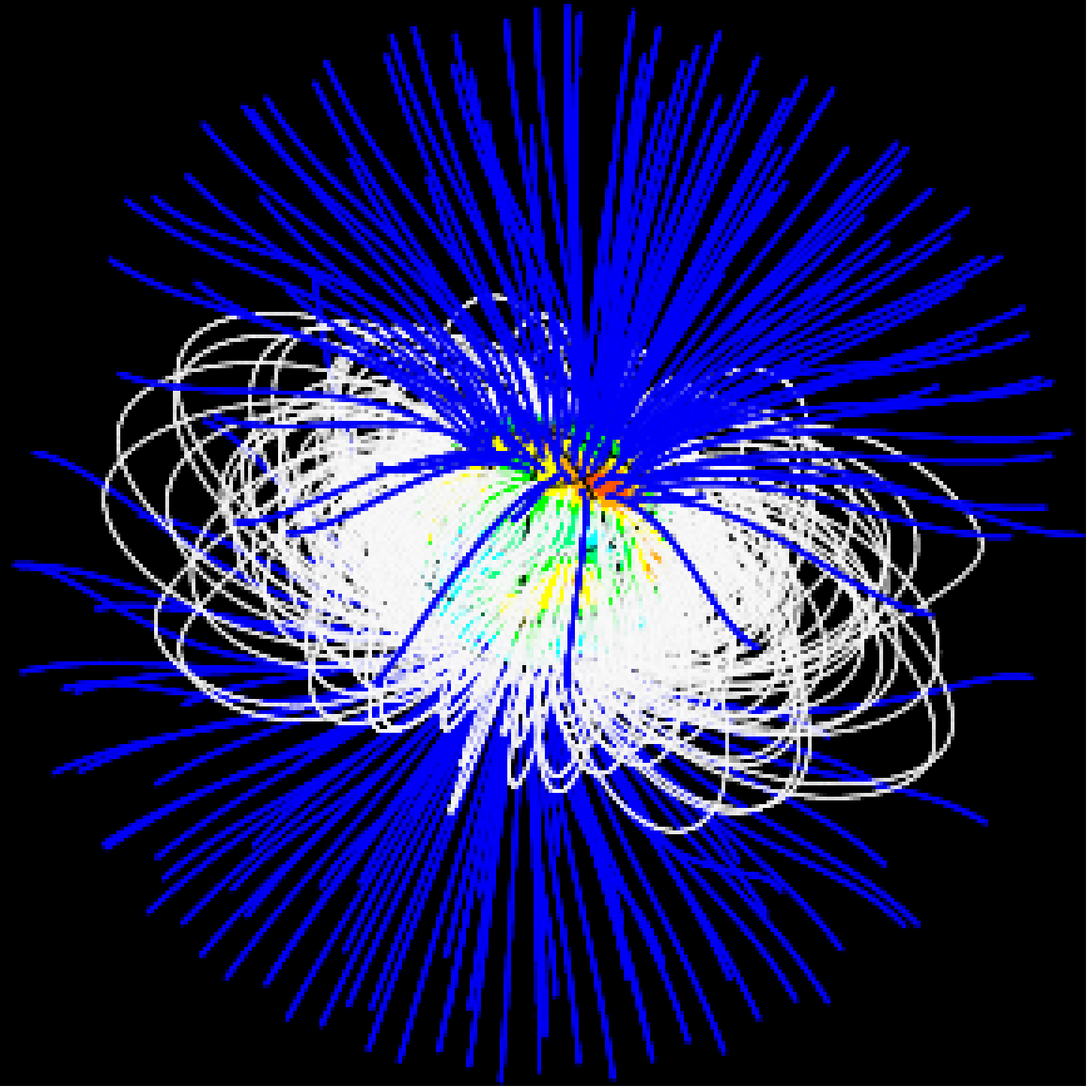}}
\caption[]{Potential extrapolations of the magnetic field reconstructed for V830~Tau in early 2016 (left) and late 2015 (right),
as seen by an Earth-based observer at phase 0.10.  Open and closed field lines are shown in blue and white respectively, whereas colors 
at the stellar surface depict the local values (in G) of the radial field (see left panels of Fig.~\ref{fig:mapv}).
The source surface at which the field becomes radial is set at a distance of 4~\rstar, close to the corotation radius of V830~Tau 
(at which the Keplerian orbital period equals the stellar rotation period and beyond which field lines tend to open under the effect 
of centrifugal forces, \citealt{Jardine04}) but smaller than the Alfv\'en radius expected for a T~Tauri star like V830~Tau 
\citep[$>$6~\rstar, see][]{Vidotto16}.  Note how the high-latitude open-field regions slightly lag behind rotation between both epochs 
as a result of differential rotation.  }
\label{fig:mag}
\end{figure*}

The large-scale magnetic topologies we retrieve for V830~Tau at both epochs (see Fig.~\ref{fig:mapv}) are again very similar, with 
rms surface magnetic fluxes of 350~G, and resemble that found previously for this star (D15).  As for the brightness maps, the main 
magnetic regions that we recover are visible at both epochs.  More specifically, the field is found to be 90\%\ poloidal, featuring a 
340~G dipole field tilted at $22\pm5$\degr\ to the rotation axis towards phase $0.79\pm0.03$ (in late 2015) and $0.88\pm0.03$ (in early 2016), 
and that gathers 60\%\ of the poloidal field energy.  
Weaker quadrupolar and octupolar components (of strength 100--150~G) and smaller-scale features are also present on V830~Tau, 
giving the field close to the stellar surface a more complex appearance than that of the dominating dipole.  With a rms flux of $\simeq$110~G, the toroidal field 
is weak and of rather complex topology.  The extrapolated large-scale magnetic topology (in the assumption of a potential field) is 
shown in Fig.~\ref{fig:mag} at both epochs.  

As for the brightness maps, the magnetic images show evidence of a global differential rotation shear similar to that reported by D15, 
with equatorial regions (e.g., the strong negative azimuthal feature at phase 0.17) moving to slightly earlier phases from late 2015 
to early 2016, and higher latitude regions (e.g., the positive radial field region at phase 0.05 and latitude 60\degr) moving to later 
phases at the same time.  The increase in the phase towards which the dipole is tilted (0.79 and 0.88 in late 2015 and early 2016 respectively) 
comes as additional evidence that high latitudes (at which the dipole poles are anchored) are rotating more slowly than average, by 
typically 1 part in 200;  this is further confirmed by the fact that the line-of-sight projected (longitudinal) magnetic fields 
\citep[proportional to the first moment of the Stokes $V$ profiles, e.g.,][and most sensitive to the low-order components of the 
large-scale field]{Donati97b} exhibit a recurrence timescale of $1.004\pm0.003$~\Prot, i.e.\ slightly longer than \Prot\ by a 
similar amount.  

We also report that the phase of maximum \hal\ emission of V830~Tau coincides, in both late 2015 and early 2016, with that of the 
high-latitude regions at which the dipole field is anchored;  this is obvious from the dynamic spectra of the \hal\ residuals that 
we provide as an additional figure in the Appendix (see Fig.~\ref{fig:haldyn}).  A logical by-product is that \hal\ emission of 
V830~Tau, like its longitudinal field, is modulated by a period slightly longer than \Prot, and equal to $1.004\pm0.002$~\Prot.  
From the solar analogy, one would have expected chromospheric emission to be minimum when open field lines point towards the observer, 
i.e., at phase 0.8--0.9 (see Fig.~\ref{fig:mag});  this is however not what we observe, suggesting that the \hal\ emission we detect 
comes from regions close to (but not coinciding with) the strongest radial field regions that we reconstruct at high latitudes (see Fig.~\ref{fig:mapv}). 

We also note apparent temporal evolution of the magnetic topology, with, e.g., the positive radial field 
region close to the equator at phase 0.33 growing much stronger between late 2015 and early 2016, though we caution again that a 
simple visual image comparison of individual features can be misleading.  

\begin{figure*}
\center{\hbox{\hspace{-2mm}
\includegraphics[scale=0.3,angle=-90]{fig/v830_drv2.ps}\hspace{2mm}
\includegraphics[scale=0.328,angle=-90]{fig/v830_dr.ps}\hspace{2mm}
\includegraphics[scale=0.3,angle=-90]{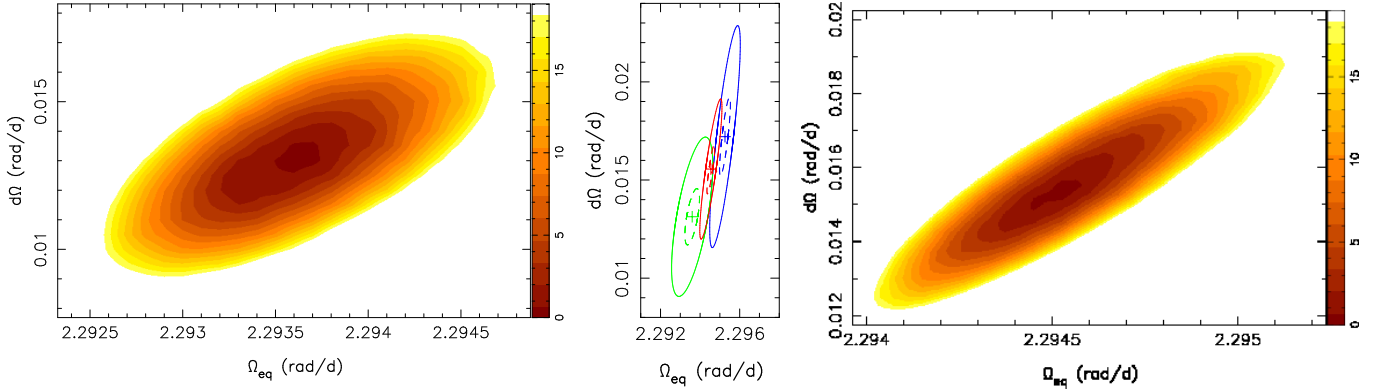}}}
\caption[]{Variations of \delchisq\ as a function of the differential rotation parameters \omeq\ and \dom, derived from modelling our 
full set of Stokes $I$ (right) and $V$ (left) LSD profiles of V830~Tau at constant information content.  
A well defined paraboloid is observed in both cases, with the outer 
color contour tracing the 99.99\% confidence interval (corresponding to a \delchisq\  of 18.4 for the 2438 Stokes $V$ and 3312 
Stokes $I$ data points).  In both cases, the minimum \chisqr\ achieved (equal to 1.62 and 1.18 for Stokes $I$ and Stokes $V$ data 
respectively) is significantly larger than 1;  this value is used to normalise \chisq\ before computing \delchisq\ so as to account 
for intrinsic variability (affecting the brightness distribution and magnetic field of V830~Tau over the course of our 91-d run, see 
Sec~\ref{sec:dr}) when estimating errors bars on differential rotation parameters.  The values we obtain for these parameters are equal to 
$\omeq=2.29455\pm0.00014$~\rpd\ and $\dom=0.0156\pm0.0009$~\rpd\ for the Stokes $I$ data, and 
$\omeq=2.29360\pm0.00025$~\rpd\ and $\dom=0.0131\pm0.0010$~\rpd\ for the Stokes $V$ data.  
The middle plot emphasizes how the confidence intervals from both measurements compare with each other, 
and with that derived from the late-2015 Stokes $I$ LSD profiles only (D16).  The 68.3\% and 99.99\% confidence intervals 
(dashed and full lines) are shown in green, red and blue for the full Stokes $V$, the full Stokes $I$ and the late-2015 Stokes $I$ 
data sets respectively. }
\label{fig:dr}
\end{figure*}

\subsection{Intrinsic variability and surface differential rotation}
\label{sec:dr}

The most reliable way to assess whether intrinsic variability occurred at the surface of V830~Tau between late 2015 and early 2016 is to 
attempt modelling both data sets simultaneously with a unique brightness and magnetic topology, and see whether one can fit the full set to the 
same \chisqr\ level as that achieved for the individual sets (i.e., 1.0, see Sec.~\ref{sec:mod}).  We find that this is not possible, 
with a minimum achievable \chisqr\ of 1.62 and 1.18 for Stokes $I$ and Stokes $V$ data respectively (starting from initial \chisqr\ of 35 
and 5);  this confirms our previous suspicion that intrinsic variability occurred at the surface of V830~Tau throughout the 91~d (33 
rotation cycles) of our observing campaign, and in particular over the 49~d shift between our two data sets.  
The global fit to the full data set we obtain nonetheless captures most of the observed line profile 
fluctuations, indicating that the intrinsic variability at work at the surface of V830~Tau remained moderate and local without altering 
the brightness and magnetic surface distributions too drastically;  this further confirms our visual impression that images from both 
epochs shared obvious similarities.  

Despite this intrinsic variability, we attempted to estimate differential rotation from our full data set.  As in previous papers, we 
achieve this by assuming that the rotation rate at the surface of V830~Tau $\Omega(\theta)$ varies with latitude $\theta$ as $\sin^2 \theta$ 
and depends on 2 main parameters, the rotation rate at the equator \omeq\ and the difference in rotation rate \dom\ between the equator and 
the pole (so that $\Omega(\theta)=\omeq-\dom\sin^2\theta$).  
Both parameters are derived by looking for the pair that minimizes the \chisqr\ of the fit to the data (at constant 
information content in the reconstructed image), whereas the corresponding error bars are computed from the curvature of the \delchisq\ 
paraboloid at its minimum \citep{Donati03b}.  (\delchisq\ is defined as the \chisq\ increase with respect to the minimum \chisq\ in the 
map.)  Results are shown in Fig.~\ref{fig:dr}.  The differential rotation we derive from our 
complete data set is slightly smaller (though still compatible at a $\simeq$3$\sigma$ level) than that inferred from the late-2015 
Stokes $I$ LSD profiles only (D16).  Despite the fact that this weakening is observed in both Stokes $I$ and $V$ data, 
we think that this small change likely results from intrinsic variability at the surface of V830~Tau\footnote{For this reason, 
the differential rotation parameters of D16 were used as reference throughout this paper, their impact on most results being however 
quite small given how weakly the photosphere of V830~Tau is sheared.}.  

Further evidence that high latitudes of V830~Tau are rotating more slowly than average (in agreement with the differential rotation pattern 
we recover) comes from the drift to later phases of the polar regions at which the large-scale dipole field component is anchored and 
where \hal\ emission is strongest.

\section{Filtering the activity jitter and modelling the planet signal}
\label{sec:fil}

We describe below the results of 3 independent techniques aimed at characterizing the RV signature of V830~Tau~b from our data.  
The first 2 methods are those already outlined in D16 and used to detect V830~Tau~b from the late-2015 data alone, that we now apply 
to both late-2015 and early-2016 data sets, with some modifications to account for the intrinsic variability between the 2 epochs (see Sec.~\ref{sec:dr}).  
The third one follows the approach of \citet{Haywood14} and \citet{Rajpaul15}, and uses Gaussian-process regression (GPR) to model 
activity directly from the raw RVs.  The results obtained with each technique are described and compared in the following sections.  

\subsection{Modeling the planet signal from filtered RVs (ZDI~\#1)}

The first technique consists in using the ZDI brightness images of Fig.~\ref{fig:mapi} to predict the RV curves expected for 
V830~Tau at each epoch, and compare them with observed raw RVs.  Modeled and raw RVs are both computed as the first order moment of 
Stokes $I$ LSD profiles (i.e., $\int (1-I(v)) v dv / \int (1-I(v)) dv$ where $v$ is the radial velocity across the line profile) 
while error bars on raw RVs are derived from those propagated from the observed spectra to the Stokes $I$ LSD profiles 
(and checked for consistency through simulated data sets as in D16);  
activity-filtered RVs are then derived by simply subtracting the modelled RVs from the observed ones 
(D16, see Table~\ref{tab:log}).  Even though the intrinsic variability observed at the surface of V830~Tau is only moderate (see 
Sec.~\ref{sec:dr}), using a specific ZDI map for each data subset (i.e., late 2015 and early 2016) is essential to obtain precise 
filtered RVs;  using a single image for both subsets and ignoring the temporal evolution of the surface brightness distribution 
between the two epochs (beyond that caused by differential rotation) significantly degrades the quality of the modelling and therefore 
the precision of the filtered RVs.  

\begin{figure*}
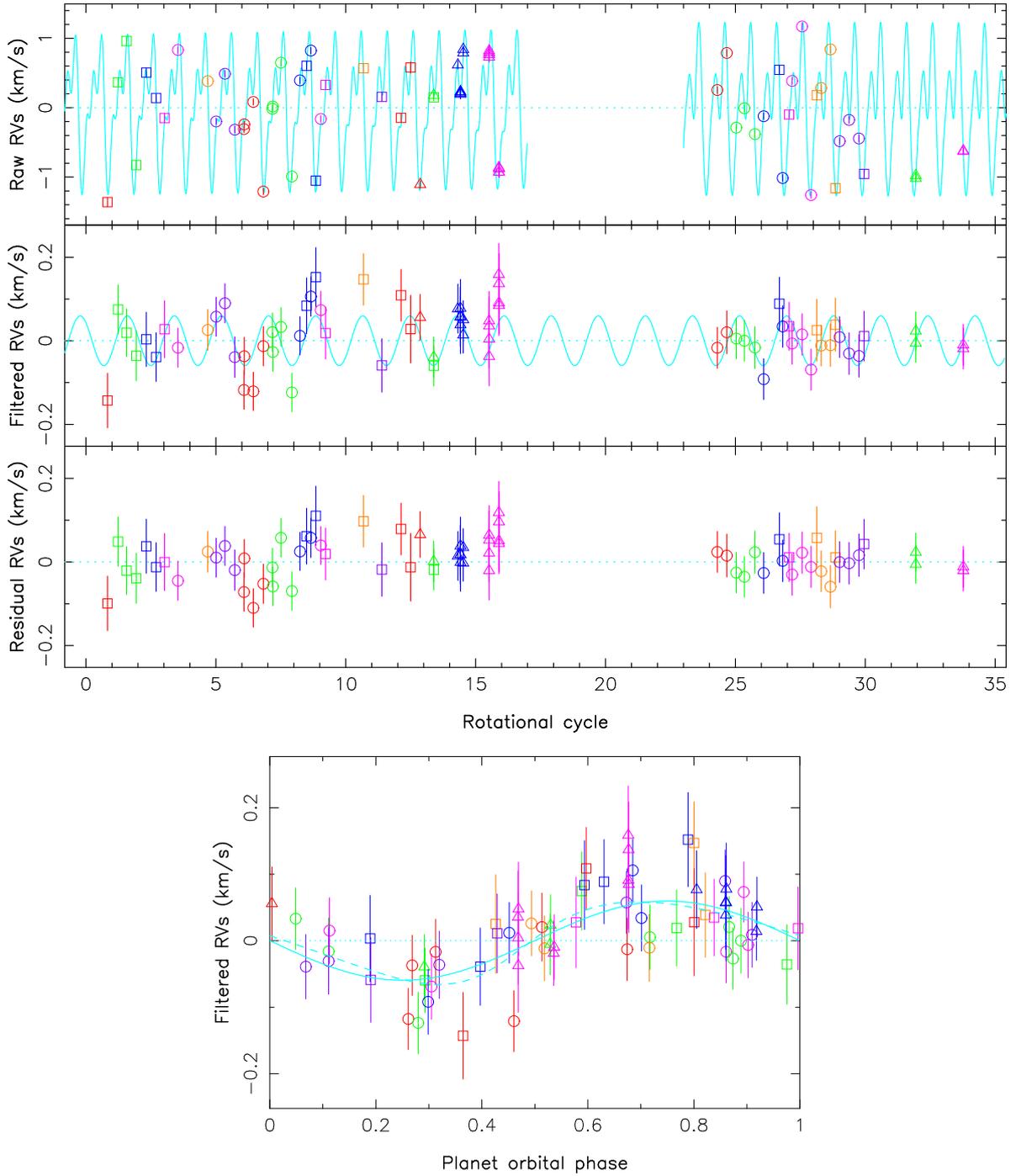

\center{
\includegraphics[scale=0.6,angle=-90]{fig/v830_sinfit.ps}\vspace{3mm}
\includegraphics[scale=0.4,angle=-90]{fig/v830_kepfit.ps}}
\caption[]{{\bf Top panel:} Raw (top), filtered (middle) and residual (bottom) RVs of V830~Tau (open symbols and 1$\sigma$ error bars, 
with circles, squares and triangles depicting ESPaDOnS, NARVAL and ESPaDOnS/GRACES data, and colors coding rotation cycles).  The raw 
RVs exhibit a semi-amplitude of 1.2~\kms\ and a rms dispersion of 0.65~\kms\ throughout the whole run.  (Rotation cycles 
of the 2016 data are shifted by $+24$ in this plot with respect to their values in Table~\ref{tab:log} and Fig.~\ref{fig:fit16}).  
The RV jitter predicted by ZDI at both epochs, as well as the best sine fit to the filtered RVs, are added in the top and middle plots 
(cyan lines).  Note how the jitter model changes between late 2015 and early 2016, and how both of them slowly evolve with time as a 
result of differential rotation.  
The rms dispersion of the residual RVs is 44~\ms, in agreement with our measurement errors (see Table~\ref{tab:log}). 
{\bf Bottom panel:} Activity-filtered RVs phase-folded on the planet orbital period.  The fit 
to the data is only marginally better with an eccentric orbit (dashed line) than with a circular one (solid line).} 
\label{fig:rvs}
\end{figure*}

\begin{figure*}
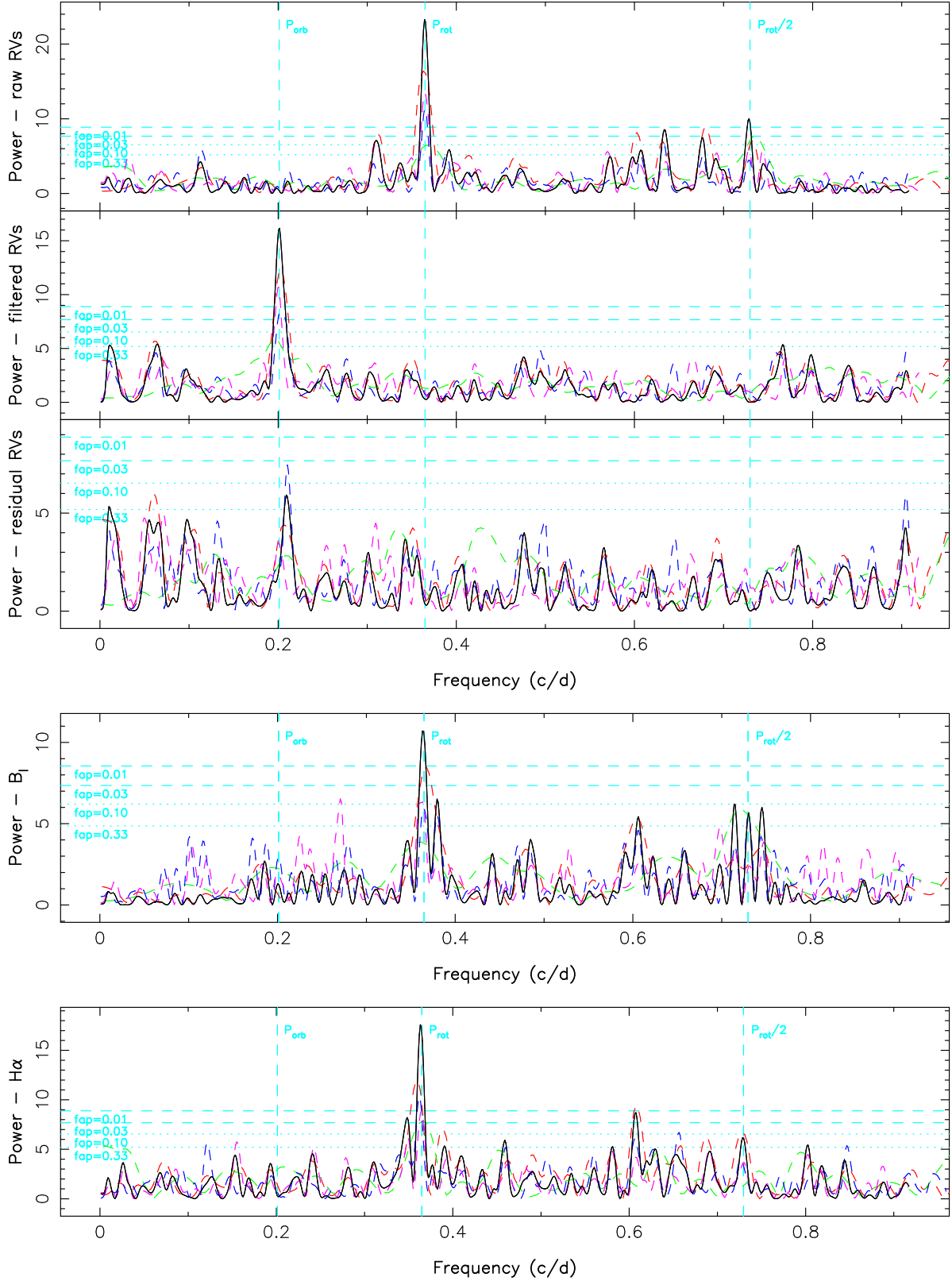

\center{
\includegraphics[scale=0.6,angle=-90]{fig/v830_mlsper.ps}\vspace{3mm}
\includegraphics[scale=0.6,angle=-90]{fig/v830_bloper.ps}\vspace{3mm}
\includegraphics[scale=0.6,angle=-90]{fig/v830_halper.ps}}
\caption[]{{\bf Top panel:} Lomb-Scargle periodograms of the raw (top), filtered (middle) and residual RVs (bottom) shown in Fig~\ref{fig:rvs}. 
The black line is for the full set, while the dashed red, green, blue and pink lines are for the late-2015, the early-2016, 
the even, and the odd points only.  The stellar rotation period, its first harmonic and the planet orbital period are depicted with 
vertical dashed lines. The horizontal dotted and dashed lines trace the 33\%, 10\%, 3\% and 1\% FAP levels.  The planet signal in the 
filtered RVs is detected in the full set with a FAP level $<10^{-5}$. {\bf Middle panel:} Periodogram of the longitudinal magnetic field, 
a reliable activity proxy \citep{Haywood16}, featuring a clear peak at the stellar rotation period but no power at the planet 
orbital period. {\bf Bottom panel:} Same as middle panel for the \hal\ emission.  }
\label{fig:pdg}
\end{figure*}

The results we obtain are shown in Fig.~\ref{fig:rvs} for the raw, filtered and residual RVs, and in Fig.~\ref{fig:pdg} for the 
corresponding periodograms.  The planet RV signal is very clearly detected in the filtered RVs, with a false-alarm probability (FAP) 
lower than $10^{-5}$.  The \chisq\ decrease that we obtain with our fit to the filtered RVs (with respect to a case with no planet) 
is about 36 (for 72 RV points and 4 degrees of freedom), suggesting a similarly-low FAP value of $<$10$^{-6}$.  
The corresponding curve features a semi-amplitude equal to $K=60\pm10$~\ms\ and an orbital period of $\Porb=4.97\pm0.03$~d, in agreement 
with the estimates of D16 ($K=75\pm12$~\ms\ and $\Porb=4.93\pm0.05$~d).  Fitting a Keplerian orbit through the data marginally improves the fit, 
but the derived eccentricity ($0.21\pm0.15$) is not measured with enough precision to be reliable \citep{Lucy71};  it confirms at least 
that V830~Tau~b is close to circular or only weakly eccentric.  The residual RVs show a rms dispersion of 44~\ms, fully compatible with 
the errors of our RV estimates (see Table~\ref{tab:log}) that mostly reflect the photon noise in our LSD profiles 
(and to a lesser extent the intrinsic RV precision of ESPaDOnS, equal to 20--30~\ms, \citealt{Moutou07, Donati08b}).  Residual RVs in 
the first part of the run (late 2015) exhibit a larger-than-average dispersion (of 50~\ms\ rms, i.e., close to the value of 48~\ms\ 
found by D16 from modelling the late 2015 data only) that mostly reflects the limits in our assumption of a constant brightness 
distribution at the surface of the star (sheared by differential rotation) on a relatively long data set (15 rotation cycles) and to a 
small extent potential residual pollution by the moon between rotational cycles 6.0 and 7.2 (see Fig.~\ref{fig:fit15}).  

Lomb-Scargle periodograms of the longitudinal fields and the \hal\ emission fluxes of V830~Tau (see Fig.~\ref{fig:pdg}, middle and bottom panels) 
both show that activity concentrates mostly at the rotation period (with a recurrence period slightly longer than \Prot, see 
Sec.~\ref{sec:mod}) and first harmonic, but not in a significant way at the planet orbital period.  This further confirms that the 
RV signal from V830~Tau~b cannot be attributed to activity.  

\subsection{Deriving planet parameters from LSD Stokes $I$ profiles (ZDI~\#2)}

The second method, proposed by \citet{Petit15} and inspired from our differential rotation measurement technique, directly works 
with Stokes $I$ LSD profiles, and consists in finding out the planet characteristics and brightness distribution that best explain the 
observed profile modulation.  More specifically, we assume the presence of a close-in planet in circular orbit with given parameters 
($K$, \Porb\ and phase of inferior conjunction), correct our LSD profiles from the reflex motion induced by the planet, reconstruct with ZDI the 
brightness image associated with the corrected LSD profiles at given information content (i.e., image spottedness) and iteratively derive 
which planet parameters allow the best fit to the data.  This technique was found to yield results in agreement with those our first 
direct method gave when previously applied to our V830~Tau data (D15, D16).  

The method was slightly modified to handle 2 different subsets of data at the same time, following \citet{Yu16}.  The main 
difference is that, for each set of planet parameters, we now reconstruct 2 different brightness images (one for each subset) with ZDI, 
both at constant information content;  we then compute a global \chisqr\ for this dual image reconstruction as a weighted mean of the 
\chisqr's associated with the 2 ZDI images (with weights equal to the number of data points in the subsets).  This allows us in particular 
to handle different brightness distributions for different epochs, without which data cannot be optimally fitted as a result of the 
intrinsic variability that the spot configuration is subject to (see Sec.~\ref{sec:dr}).  

\begin{figure}
\includegraphics[scale=0.35,angle=-90]{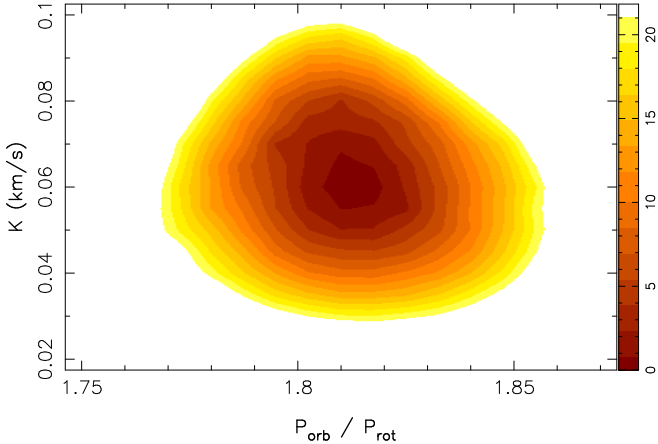}
\caption[]{Variations of \delchisq\ of the ZDI fits to the late-2015 and early-2016 LSD profiles of V830~Tau (for a fixed spottedness 
level at both epochs), after removing the reflex motion of a close-in a planet and for a range of orbital periods \Porb\ (actually the 
ratio of the orbital to rotation period $\Porb/\Prot$) and semi-amplitudes $K$ of the planet RV signature. (This is a 2D cut from a 3D map, 
with the phase of the RV signal also included as a search parameter).  A clear minimum is obtained in the \delchisq\ landscape, whose 
projection in a $K$ vs $\Porb/\Prot$ plane passing through the minimum is shown here.  The outer color contour traces the projected 99.99\% 
confidence interval, corresponding to a \delchisq\ of 21.1 for a 3-parameter fit to the 3312 data points of the LSD profiles. } 
\label{fig:ppar}
\end{figure}

The planet parameters we derive with this second technique are equal to $K=62\pm9$~\ms\ and $\Porb=4.97\pm0.03$~d, very similar to those 
obtained with our first method and again in agreement with those of D16.  The corresponding \chisq\ map (projected onto the $K$ vs $\Porb$ 
plane that passes through the global minimum), shown in Fig.~\ref{fig:ppar}, features a clear minimum.  With respect to our best model 
incorporating a planet, a model with no planet corresponds to a \delchisq\ of 75, indicating that the planet is detected with a 
FAP level $<10^{-15}$;  the much lower FAP directly reflects the larger \delchisq\ obtained with this method, reflecting that 
line profiles of rapid rotators contain more (or less-noisy) information than their first moments (the raw and filtered RVs).  

\subsection{Deriving planet parameters from raw RVs using Gaussian-process regression (GPR)}

The third method we applied to our data works directly from raw RVs and uses GPR to model the activity jitter as well 
as its temporal evolution, given its covariance function \citep[e.g.,][]{Haywood14, Rajpaul15}.  Assuming again the presence of a 
close-in planet of given characteristics, we correct the raw RVs from the reflex motion induced by the planet and fit the 
corrected RVs with a Gaussian process (GP) based on a pseudo-periodic covariance function $c(t,t')$ of the form: 
\begin{eqnarray}
c(t,t') & = & \theta_1^2 \exp \left( -\frac{(t-t')^2}{\theta_3^2} -\frac{\sin^2 \left( \frac{\pi (t-t')}{\theta_2} \right)}{\theta_4^2} \right) 
\label{eq:covar}
\end{eqnarray}
where $\theta_1$ is the amplitude of the GP (in \kms), $\theta_2$ the recurrence timescale (i.e., close to 1 here, in units of \Prot), 
$\theta_3$ the decay timescale (i.e., the typical spot lifetime here, in units of \Prot) and $\theta_4$ a smoothing parameter (within [0,1]) 
setting the amount of high frequency structure that we allow the fit to include.  For a given set of planet parameters and of the 4 GP hyper 
parameters $\theta_1$ to $\theta_4$, we can compute the GP that best fits the corrected raw RVs (denoted $y$) and estimate the log likelihood 
$\log \mathcal{L}$ of the corresponding parameter set from: 
\begin{eqnarray}
2 \log \mathcal{L} & = & -n \log(2\pi) - \log|C+\Sigma| - y^T (C+\Sigma)^{-1} y
\label{eq:llik}
\end{eqnarray}
where $C$ is the covariance matrix for all observing epochs, $\Sigma$ the diagonal variance matrix of the raw RVs and $n$ the number of data points.  Coupling this with 
a Markov Chain Monte-Carlo (MCMC) simulation to explore the parameter domain, we can determine the optimal set of planet and GP hyper parameters 
that maximizes likelihood, as well as the relative probability of this optimal model with respect to one with no planet (and only the GP modelling activity).  

We start by carrying out an initial MCMC run with input priors, whose results (the posterior distributions) are used to infer refined priors and 
proposal distributions capable of ensuring both an efficient mixing and convergence of the chain as well as a thorough exploration of the domain 
of interest (through a standard Metropolis-Hastings jumping scheme);  these refined priors are found to be weakly dependent on the input priors, 
already suggesting that our data contain enough information to reliably characterize the GP and planet parameters.  The main MCMC run uses 
our refined priors, listed in Table~\ref{tab:gpp} for the various parameters;  we usually carry out two successive main runs, a first one with all 
4 GP hyper parameters and 3 planet parameters free to vary, then a second one with both $\theta_3$ and $\theta_4$ fixed to their best values and 
the remaining 5 parameters left free to vary.  The goal of this sequential approach is to incorporate as much prior information about the stellar 
activity as possible into our model (hence the stronger refined priors) so that the GP yields a robust estimation of the uncertainties on the 
final parameters (particularly the planet mass) given these priors (see \citealt{Haywood14} and \citealt{Lopez16} for a similar approach).  

\begin{table}
\caption[]{Priors used in our MCMC simulation for the planet and GP hyper parameters.  
We mention the mean and standard deviation used for the refined Gaussian priors (plus the standard 
deviation assumed for our initial MCMC run), the minimum and maximum 
values allowed for the uniform and Jeffreys priors, as well as the knee value for the modified 
Jeffreys priors (following \citealt{Haywood14}, with $\sigma_{\rm RV}$ noting the $\sn^2$-weighted 
average RV error of our measurements, equal to 53~\ms, see Table~\ref{tab:log}).  The planet phase 
$\phi_0$ relates to the epoch of inferior conjunction \tconj\ through $\tconj=\phi_0\Porb + t_0$ 
where $t_0=2,457,359.9069$~d (corresponding to rotation cycle 127.0) is the reference zero time we 
used for our observations.  } 
\center{
\begin{tabular}{cc}
\hline
Parameter   & Prior   \\ 
\hline
\Porb/\Prot & Gaussian (1.80, 0.012, initial 0.10) \\
$K$ (\kms)  & modified Jeffreys ($\sigma_{\rm RV}$)\\
$\phi_0$    & Gaussian (0.13, 0.04, initial 0.10) \\
\hline
GP amplitude        $\theta_1$ (\kms)  & modified Jeffreys ($\sigma_{\rm RV}$) \\
Recurrence period   $\theta_2$ (\Prot) & Gaussian (1.0, 0.001, initial 0.010) \\
Spot lifetime       $\theta_3$ (\Prot) & Jeffreys (0.1, 500.0) \\
Smoothing parameter $\theta_4$         & Uniform  (0, 1) \\
\hline
\end{tabular}}
\label{tab:gpp}
\end{table}

\begin{figure*}
\includegraphics[scale=0.6,angle=-90]{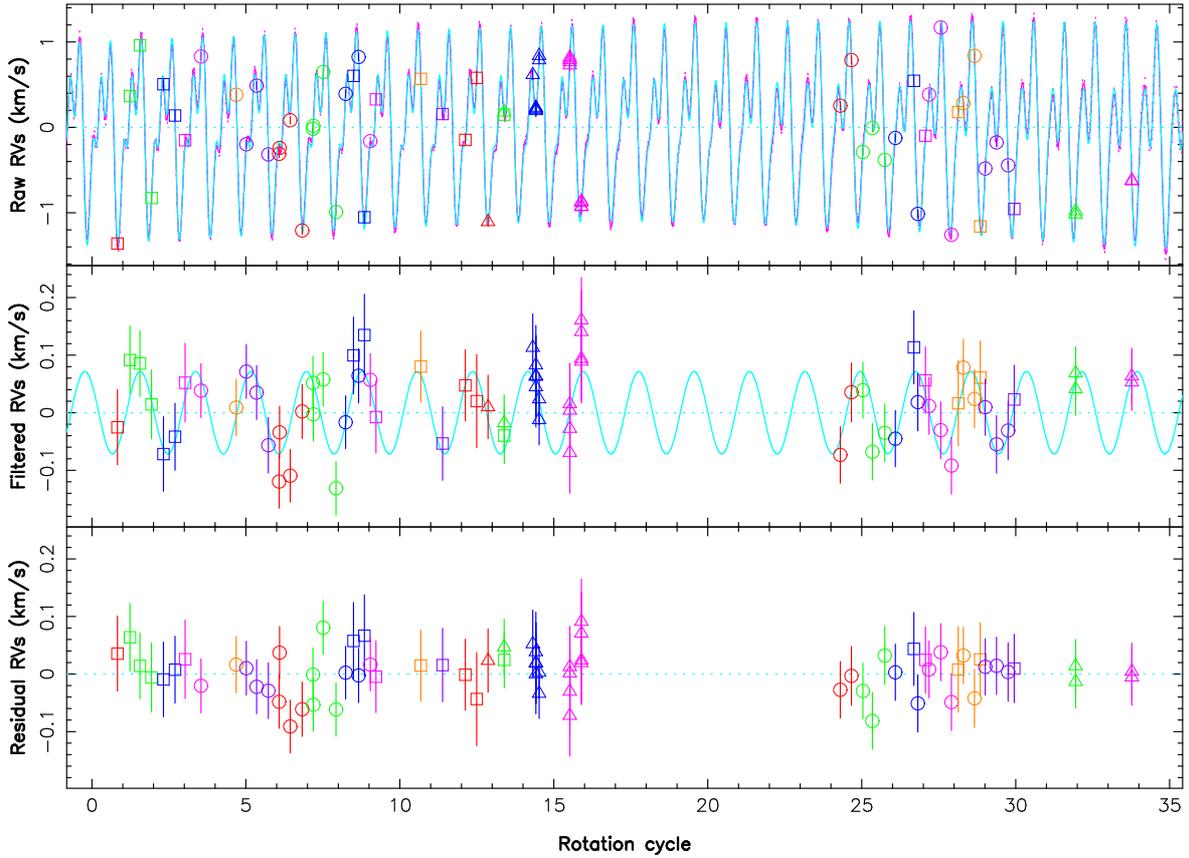}
\caption[]{Same as Fig.~\ref{fig:rvs} but now using a planet+GP fit to the data (pink line), where the GP is modelling the activity jitter (cyan line) 
while the planet and GP parameters are determined using a MCMC simulation.  Note how the GP succeeds at modelling the activity and its temporal evolution 
throughout the whole observing window, and not just for the two separate subsets. The rms dispersion of the residual RVs is 37~\ms.  } 
\label{fig:gpfit}
\end{figure*}

We find that $\theta_3$, the hyper parameter describing spot lifetime, gives the best result for a value of $\theta_3=44\pm11~\Prot = 120\pm30$~d, 
only slightly longer than the full duration of our observing run (91~d).  This further confirms the importance of taking into account 
the temporal evolution of brightness maps in activity filtering studies, even in the case of wTTSs like V830~Tau whose spot distributions 
are known to be fairly stable on long timescales;  whereas this is true for the largest surface features, this is no longer the case for 
the smaller ones whose effect on RV curves is significant.  Similarly, we get that $\theta_4=0.6\pm0.1$ yields the most likely fit to the data;  
this reflects the lack of fine structure in the RV curves, as expected from the fact that RVs are the first-order moment of Stokes $I$ LSD 
profiles that acts as a low-pass filter on surface brightness distributions.  
With the final MCMC run, we obtain that the recurrence timescale $\theta_2$ is equal to $\theta_2=0.9986\pm0.0007~\Prot$, i.e., only very 
slightly shorter than the average rotation period $\Prot$ on which our data were phased (see Eq.~\ref{eq:eph});  we note that this period matches well 
the equatorial rotation period of V830~Tau (see Table~\ref{tab:spar} and Sec.~\ref{sec:dr}), suggesting that RVs are primarily affected by 
equatorial features at the stellar surface.  For the GP amplitude 
$\theta_1$, we find that $\theta_1=0.878\pm0.135$~\kms, $\simeq$30\% larger than the rms dispersion of our raw RVs (equal to 0.65~\kms\ prior to 
any activity filtering, or removal of planetary-induced reflex motions).  

For the planet parameters, we find that $K=68\pm11$~\ms\ and $\Porb=4.93\pm0.03$~d, whereas the most accurate epoch of inferior conjunction 
(assuming a circular orbit) is found to be $\tconj=2,457,360.51\pm0.14$. The corresponding fit to the data, shown in 
Fig.~\ref{fig:gpfit}, 
demonstrates that the GP is doing a very nice job at modelling not only the activity, but also its evolution with time.  Comparing with 
the results of our first method (see Fig.~\ref{fig:rvs}), we can see that both the GP and ZDI predict similar RV curves.  However, 
thanks to its higher flexibility, the GP does a better job at matching the data, not only for our second data set where temporal 
variability is higher (given the faster evolution of the predicted RV curve, see Fig.~\ref{fig:gpfit}) and where the planet signal is 
clearly better recovered, but also for our first data set where the slower spot evolution is enhanced by the longer time span (of 15 rotation 
cycles).  As a result, the rms dispersion of the RV residuals has further decreased to 37~\ms, 16\% smaller than with our first method, 
including in the first part of our run (late 2015) where the fit to the data is now tighter (rms dispersion of RV residuals of 40~\ms\ 
instead of 50~\ms, and close to that of the full run).  
Given this, we consider that the planet parameters derived with this third method, and in particular $K$ and \Porb, 
are likely more accurate than the estimates obtained with the two previous techniques;  they also agree better with the initial estimates 
of D16 inferred from the late 2015 data only.  The phase plots of our
final 5-parameter MCMC run are provided in Appendix~\ref{sec:fig} (see Fig.~\ref{fig:mcmc}, left panel), showing little correlation 
between the various parameters and thus minimum bias in the derived values.  

When applying this technique to the full series of raw RVs collected to date on V830~Tau, including our original set secured in 
late 2014 and early 2015 (D15, D16), we further enhance the precision 
on the derived parameters, in particular on the orbital period that we can now pin down to $\Porb=4.927\pm0.008$~d.  The derived 
semi-amplitude of the RV curve is the same as in the previous fit ($K=68\pm11$~\ms) whereas the epoch of inferior conjunction
(assuming a circular orbit) is only slightly improved ($\tconj=2,457,360.522\pm0.124$).  The phase plots of this MCMC run are 
also provided in Appendix~\ref{sec:fig} (see Fig.~\ref{fig:mcmc}, right panel).  

Applying the method of \citet{Chib01} to the MCMC posterior samples, we obtain that the marginal likelihood of the model including 
the planet is higher than that of a model with no planet by a Bayes' factor of $10^8$ ($10^9$ when also including our raw RVs 
from late 2014 and early 2015), providing a strong and independent confirmation that V830~Tau hosts a close-in giant planet in a 4.93~d orbit.  
Assuming now a planet on an elliptical orbit (and using $\sqrt{e} \cos \omega$ and $\sqrt{e} \sin \omega$ as search parameters where $e$ and $\omega$ 
respectively denote the eccentricity and argument of periapsis of the orbit, \citealt{Ford06}) yields a low eccentricity of $0.16\pm0.20$;  
the marginal likelihood of this latter model is however larger than that of the circular planet model by a Bayes' factor of $<3$, 
implying that there is no evidence yet that the planet is eccentric.  We provide the MCMC phase plots of the eccentric orbit model in Fig.~\ref{fig:mcmc2}.  

The planet parameters derived with all 3 methods are summarized in Table~\ref{tab:tpar}, with those derived in D16 (from our 
late 2015 data only) listed as well for an easy comparison.  

\begin{table*}
\caption[]{Summary of our results using the three filtering techniques described in Sec.~\ref{sec:fil};  columns~2 to 4 are 
for the 2 ZDI-based and GPR methods described in Secs.~4.1 to 4.3 and applied to our new data from late 2015 to early 2016, 
whereas column~5 is for GPR applied to the entire data set (including that of D15).  
Column~6 recalls the results derived in D16 from the late~2015 data only as a comparison.  
The first table section lists the derived planet parameters (with $M_{\rm p}$ denoting the planet mass), the second one 
mentions the inferred GP hyper parameters in the GPR case, and the last one recalls the achieved \chisqr\ (to the filtered 
RVs for ZDI \#1, to the Stokes $I$ profiles for ZDI \#2 and to the raw RVs for GPR) and the rms dispersion of the RV residuals 
(whenever relevant). } 
\center{
\begin{tabular}{cccccc}
\hline
Parameter                  & ZDI \#1          & ZDI \#2            & GPR                & GPR (all data)   & D16 \\ 
\hline
\Porb\ (d)                 & $4.97\pm0.03$    & $4.97\pm0.03$      & $4.93\pm0.03$      & $4.927\pm0.008$  & $4.93\pm0.05$  \\ 
$K$ (\ms)                  & $60\pm10$        & $62\pm9$           & $68\pm11$          & $68\pm11$        & $75\pm12$      \\ 
$\phi_0$                   & $0.128\pm0.025$  & $0.142\pm0.024$    & $0.122\pm0.028$    & $0.125\pm0.025$  & $0.123\pm0.025$ \\ 
\tconj\ (2,457,300+)       & $60.54\pm0.13$   & $60.61\pm0.12$     & $60.51\pm0.14$     & $60.523\pm0.124$ & $60.52\pm0.13$ \\ 
$M_{\rm p} \sin i$ (\mjup) & $0.50\pm0.09$    & $0.52\pm0.09$      & $0.57\pm0.10$      & $0.57\pm0.10$    & $0.63\pm0.10$  \\ 
$M_{\rm p}$ (\mjup) assuming $i=55\degr$ & $0.61\pm0.11$ & $0.63\pm0.11$ & $0.70\pm0.12$ & $0.70\pm0.12$   & $0.77\pm0.12$  \\ 
$a$ (au)                   &                  &                    &                    & $0.057\pm0.001$  &     \\
$a/\rstar$                 &                  &                    &                    & $6.1\pm0.6$      &     \\ 
\hline
GP amplitude $\theta_1$ (\kms)       &        &                    & $0.878\pm0.135$    & $0.842\pm0.105$  &     \\ 
Recurrence period $\theta_2$ (\Prot) &        &                    & $0.9986\pm0.0007$  & $0.9985\pm0.0006$&     \\ 
Spot lifetime $\theta_3$ (d)         &        &                    & $120\pm30$         & & \\ 
Smoothing parameter $\theta_4$       &        &                    & $0.6\pm0.1$        & & \\ 
\hline
\chisqr\                   & 0.68             & 1.0                & 0.48               & 0.42             & 0.75 \\ 
rms RV residuals (\ms)     & 44               &                    & 37                 & 35               & 48  \\ 
\hline
\end{tabular}} 
\label{tab:tpar}
\end{table*}

\section{Summary \& discussion}
\label{sec:dis}

This paper reports the results of an extended spectropolarimetric run on the wTTS V830~Tau, carried out in the framework of the 
international MaTYSSE Large Programme, using ESPaDOnS on the CFHT, Narval on the TBL and GRACES/ESPaDOnS on Gemini-North, spanning from 
2015 Nov~11 to Dec~22, then from 2016 Jan~14 to Feb~10, and complemented by contemporaneous photometric observations from the 
\hbox{1.25-m} telescope at CrAO.  This new study is an in-depth follow-up of a previous one, based only on the first part of 
this data set and focussed on the detection of the young close-in hJ orbiting V830~Tau in 4.93~d (D16), and of an older one 
that suspected the presence of V830~Tau~b, but from too sparse a data set to firmly demonstrate the existence of the planet (D15).  

Applying ZDI to our two new data subsets, we derived the surface brightness and magnetic maps of V830~Tau.  Cool spots and warm 
plages are again present on V830~Tau, totalling 13\%\ of the overall stellar surface for those to which ZDI is sensitive.  The 
brightness maps from late 2015 and early 2016 are similar, except for differential rotation slightly shearing the photosphere of 
V830~Tau and for small local changes in the spot distribution, reflecting their temporal evolution on a timescale of only a few weeks.  
The magnetic maps of V830~Tau are also quite similar at both epochs and to that reconstructed from our previous data set (D15), 
featuring a mainly poloidal field whose dominant component is a 340~G dipole tilted at 22\degr\ from the rotation axis.  As for 
the brightness distribution, the magnetic field is also sheared by a weak surface differential rotation, and is evolving with 
time over the duration of our observing run.  

We detected several flares of V830~Tau during the second part of our run, where one major event and a weaker precursor 
were strong enough to impact RVs at a level of about 0.3~\kms.  In addition to generate intense emission in the usual spectral 
activity proxies including the \hal, \caii\ IRT and \hei\ $D_3$ lines, these flares triggered large redshifts of the emission 
component, especially for the \hei\ $D_3$ line whose redshift reaches up to 35~\kms\ with respect to the stellar rest frame, and 
25~\kms\ with respect to the average line position in a quiet state.  By analogy with the Sun and young active stars 
\citep[e.g.,][]{Cameron89, Cameron89b}, we propose that the flares we detect on V830~Tau relate to coronal mass ejections and 
reflect the presence of massive prominences in the magnetosphere of V830~Tau, likely confined by magnetic fields in the equatorial 
belt of closed-field loops encircling the star (see Fig.~\ref{fig:mag}), and whose stability is perturbed by the photospheric shear 
stressing the field or by the hot Jupiter itself in the case of large magnetic loops extending as far as the giant planet orbit 
(at 6.1~\rstar).  High-cadence spectral monitoring in various activity proxies is required to investigate such flares in more detail, 
work out the fate of associated prominences once no longer magnetically confined, and diagnose the main triggering mechanism behind 
them.  

We applied 3 different methods to our full data set to further confirm the existence of its hJ, and better characterize 
its orbital parameters.  The first two methods, using ZDI to model and predict the RV activity jitter, are those with which 
V830~Tau~b was originally detected, in a slightly modified version allowing them to handle two different ZDI images (corresponding 
to the late-2015 and early-2016 subsets) at the same time and account for the potential evolution of brightness distributions between 
the 2 epochs.  Our third technique is fully independent from the 2 others and directly 
works from raw RVs, using GPR to model the RV activity jitter and MCMC to infer the optimal planet and GP parameters and error 
bars in a Bayesian formalism, following \citet{Haywood14}.  All 3 methods unambiguously confirm the existence of V830~Tau~b and 
yield consistent results for the planet parameters when applied to our new data;  in particular, all are able to reliably recover 
the RV planet signal (of semi-amplitude $68\pm11$~\ms) hiding behind the activity jitter (of semi-amplitude 1.2~\kms\ and rms dispersion 
0.65~\kms) that the brightness distribution of V830~Tau is inducing.  
The third method is found to perform best, thanks to its higher flexibility and better performances at modelling the temporal 
evolution of the RV activity jitter.  Applying this third method to all raw RVs collected to date on V830~Tau (including those of 
D15) allows us to significantly improve the precision on the planet orbital period.  We also confirm that the planet orbit is more or less 
circular, with no evidence for a non-zero eccentricity at a 1$\sigma$ precision of 0.15--0.20. 
Further work is needed to enable ZDI reconstructing time-variable features and make it as efficient as GPR for filtering activity 
from RV curves of young active stars.  

Spectropolarimetry is found to be essential for retrieving the large-scale topology of the magnetic field that fuels all activity phenomena, 
but not critical for modelling and filtering the activity jitter at optical wavelengths, largely dominated by the impact of surface 
brightness features;  however, spectropolarimetry is expected to become crucial at nIR wavelengths where brightness features contribute 
less jitter and Zeeman distortions are much larger than in the optical \citep[e.g.,][]{Reiners13, Hebrard14}.  

Along with the latest reports of similar detections (or candidate detections) of young close-in giants around TTSs 
\citep[e.g.,][]{vanEyken12, Mann16, JohnsKrull16, David16, Yu16}, our result suggests that newborn hJs may be frequent, possibly 
more so than their mature equivalents around Sun-like stars \citep{Wright12}.  The orbital fate of young hJs like V830~Tau~b under 
tidal forces and strong winds as the host star progresses on its evolutionary track, contracts and spins up to the main sequence, and 
at the same time loses angular momentum to its magnetic wind and planet, is still unclear \citep[e.g.,][]{Vidotto10, Bolmont16}.  
One can expect V830~Tau~b, whose orbital period is currently longer than the stellar spin period, to be spiralling outwards, at 
least until V830~Tau is old enough to rotate more slowly than its close-in giant;  investigating whether tidal 
forces will still be strong enough by then to successfully drag V830~Tau~b back and kick it into its host star in the next few 
hundred Myrs, may tell whether and how frequent newborn close-in giants can be reconciled with the observed sparse population 
of mature hJs.  

Alternatively, the MaTYSSE sample may be somehow biased towards wTTSs hosting hJs \citep[e.g.,][]{Yu16}.  In particular, our sample is 
likely biased towards wTTSs whose discs have dissipated early, i.e., at a time where the star, still fully convective, hosted a magnetic 
field strong enough to carve a large magnetospheric gap \citep{Gregory12, Donati13} and trigger stable accretion \citep{Blinova16}.  
This may come as favorable conditions for hJs to survive type-II migration, when compared to more evolved cTTSs featuring weaker 
fields, smaller magnetospheric gaps and chaotic accretion.  

Last but not least, we stress that V830~Tau is the first known non-solar planet host that exhibits radio emission \citep{Bower16}, which 
opens very exciting perspectives for in-depth studies of star-planet interactions, and possibly even of exoplanetary magnetic fields 
\citep{Vidotto10, Vidotto16}.  

Applying the complementary detection techniques outlined in this paper to extended spectropolarimetric data sets such as those 
gathered within MaTYSSE, or forthcoming ones to be collected with SPIRou, the nIR spectropolarimeter / high-precision velocimeter 
currently in construction for CFHT (first light planned in 2017), should turn out extremely fruitful and enlightening for our 
understanding of star / planet formation, about which little observational constraints yet exist.

\section*{Acknowledgements} 
This paper is based on observations obtained at the CFHT (operated by the National Research Council of Canada / CNRC, the Institut 
National des Sciences de l'Univers / INSU of the Centre National de la Recherche Scientifique / CNRS of France and the University of 
Hawaii), at the TBL (operated by Observatoire Midi-Pyr\'en\'ees and by INSU / CNRS), and at the Gemini Observatory (operated by the 
Association of Universities for Research in Astronomy, Inc., under a cooperative agreement with the National Science Foundation / NSF 
of the United States of America on behalf of the Gemini partnership: the NSF, the CNRC, CONICYT of Chile, Ministerio de Ciencia, 
Tecnolog\'\i a e Innovaci\'on Productiva of Argentina, and Minist\'erio da Ci\^encia, Tecnologia e Inova\c c\~ao of Brazil). 
This research also uses data obtained through the Telescope Access Program (TAP), which has been funded by the National Astronomical 
Observatories of China, the Chinese Academy of Sciences (the Strategic Priority Research Program ``The Emergence of Cosmological Structures''
Grant \#XDB09000000), and the Special Fund for Astronomy from the Ministry of Finance.

We thank the QSO teams of CFHT, TBL and Gemini for their great work and efforts at collecting the high-quality MaTYSSE data presented 
here, without which this study would not have been possible.  MaTYSSE is an international collaborative research programme involving 
experts from more than 10 different countries.  

We also warmly thank the IDEX initiative at Universit\'e F\'ed\'erale Toulouse Midi-Pyr\'en\'ees (UFTMiP) for funding the STEPS 
collaboration program between IRAP/OMP and ESO and for allocating a ``Chaire d'Attractivit\'e'' to GAJH allowing her regularly visiting 
Toulouse to work on MaTYSSE data.  We acknowledge funding from the LabEx OSUG@2020 that allowed purchasing the ProLine PL230 CCD imaging 
system installed on the 1.25-m telescope at CrAO.  SGG acknowledges support from the Science \& Technology Facilities Council (STFC) 
via an Ernest Rutherford Fellowship [ST/J003255/1].  SHPA acknowledges financial support from CNPq, CAPES and Fapemig.  

We finally thank the referee, Teruyuki Hirano, for his valuable comments that helped us improve the paper.

\appendix

\section{Additional figures}
\label{sec:fig}

\begin{figure*}
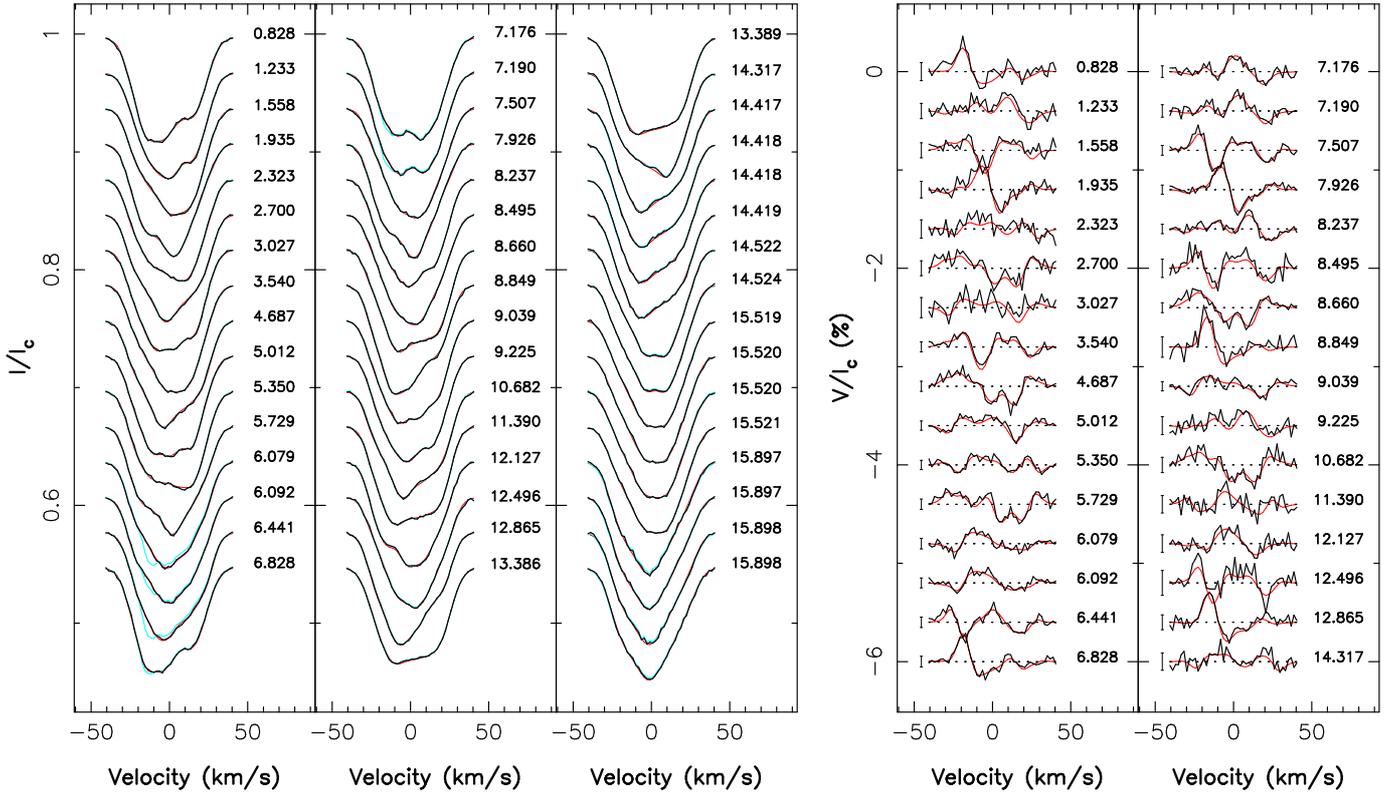

\center{
\hbox{\hspace{-3mm}
\includegraphics[scale=0.6,angle=-90]{fig/v830_fiti15.ps}\hspace{3mm}
\includegraphics[scale=0.6,angle=-90]{fig/v830_fitv15.ps}}} 
\caption[]{Same as Fig.~\ref{fig:fit16} for our late 2015 observations. } 
\label{fig:fit15}
\end{figure*}

\begin{figure*}
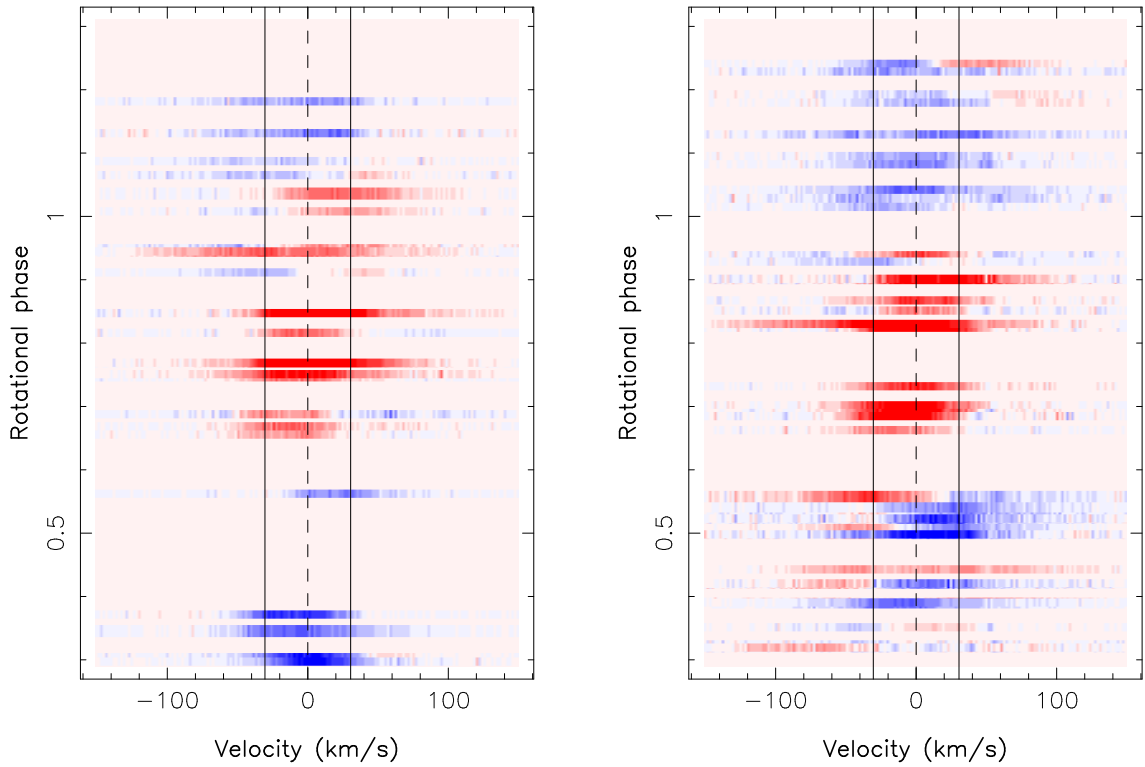

\center{
\hbox{\hspace{10mm}
\includegraphics[scale=0.43]{fig/v830_dyn16.ps}\hspace{10mm}
\includegraphics[scale=0.43]{fig/v830_dyn15.ps}}}
\caption[]{Dynamic spectra of \hal\ residual of V830~Tau in late 2016 (left) and early 2015 (right), with residuals computed with  
respect to the $\sn^2$-weighted mean over the whole observing run (after the removal of a few flaring spectra, see Sec.~\ref{sec:obs}).  
Note how the phase of maximum \hal\ emission increases from 0.8 to 0.9 from late 2015 to early 2016.  Red / blue means positive / 
negative residuals, with amplitudes ranging from --0.3 to 0.3 (in units of the continuum level), whereas the dashed and full vertical 
lines depict the line centre (in the stellar rest frame) and the stellar rotational broadening \vsini.  } 
\label{fig:haldyn}
\end{figure*}

\begin{figure*}
\hbox{
\includegraphics[scale=0.51,angle=-90]{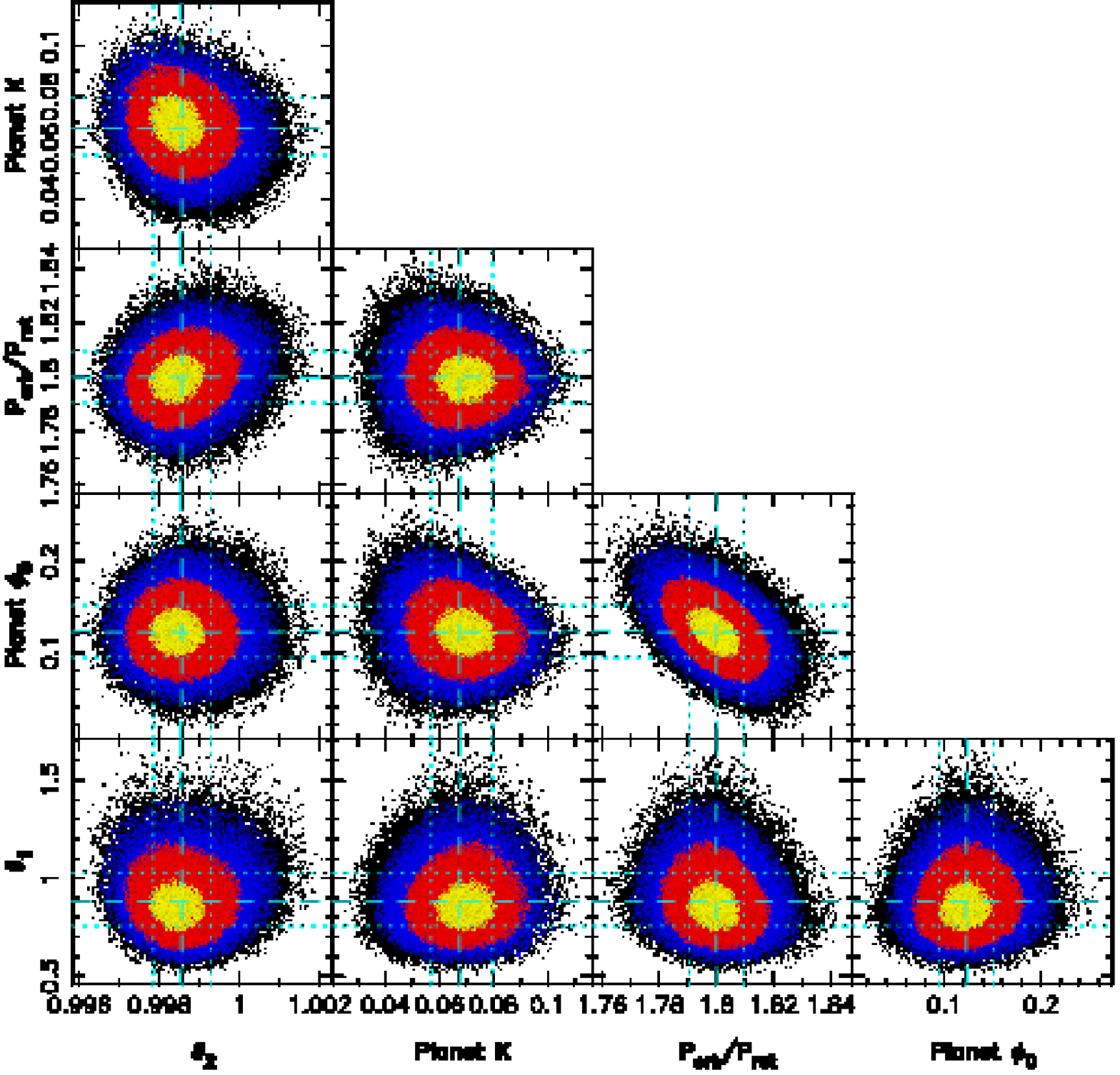}\hspace{3mm}
\includegraphics[scale=0.51,angle=-90]{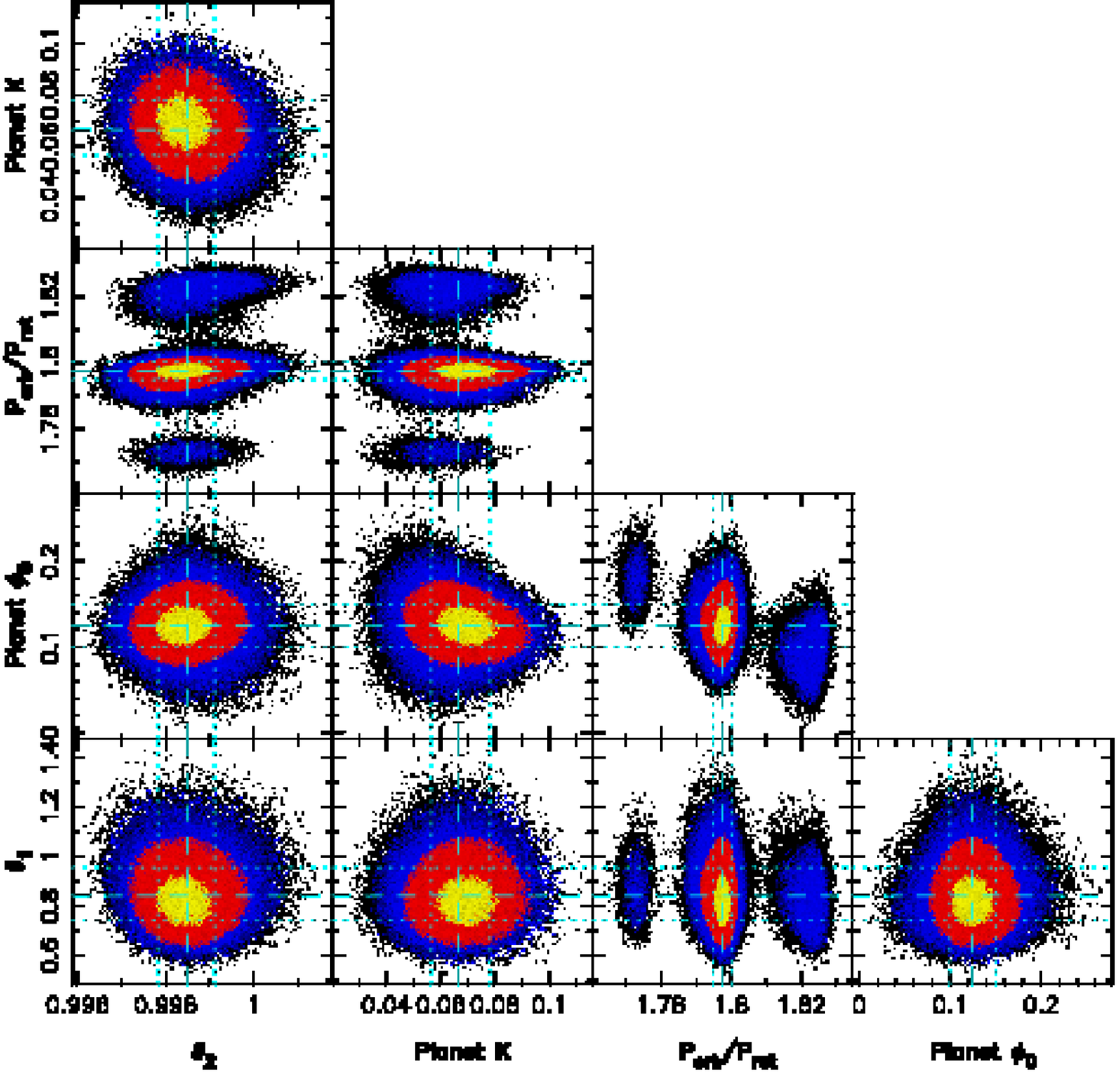}}
\caption[]{Phase plots of our final 5-parameter MCMC run using both the late 2015 and early 2016 data (left) and 
all raw RVs collected to date on V830~Tau (right), with yellow, red and blue points marking the 1, 2 and 3-$\sigma$ 
confidence regions respectively.  
The optimal parameter we derive from the left panel are respectively equal to $\theta_1=0.878\pm0.135$~\kms, 
$\theta_2=0.9986\pm0.0007~\Prot$, $K=68\pm11$~\ms, $\Porb/\Prot=1.80\pm0.01$ (i.e., $\Porb=4.93\pm0.03$~d) and 
$\phi_0=0.122\pm0.028$ (i.e., $\tconj=2,457,360.51\pm0.14$~d).  Fitting all raw RVs allows to significantly improve 
the precision on the rotation period ($\Porb/\Prot=1.7976\pm0.0027\pm0.01$, i.e., $\Porb=4.927\pm0.008$~d) and to 
slightly refine the epoch of inferior conjunction ($\phi_0=0.125\pm0.025$, i.e., $\tconj=2,457,360.523\pm0.124$~d). 
Note the little correlation between the various parameters.  } 
\label{fig:mcmc}
\end{figure*}

\begin{figure*}
\hbox{
\includegraphics[scale=0.51,angle=-90]{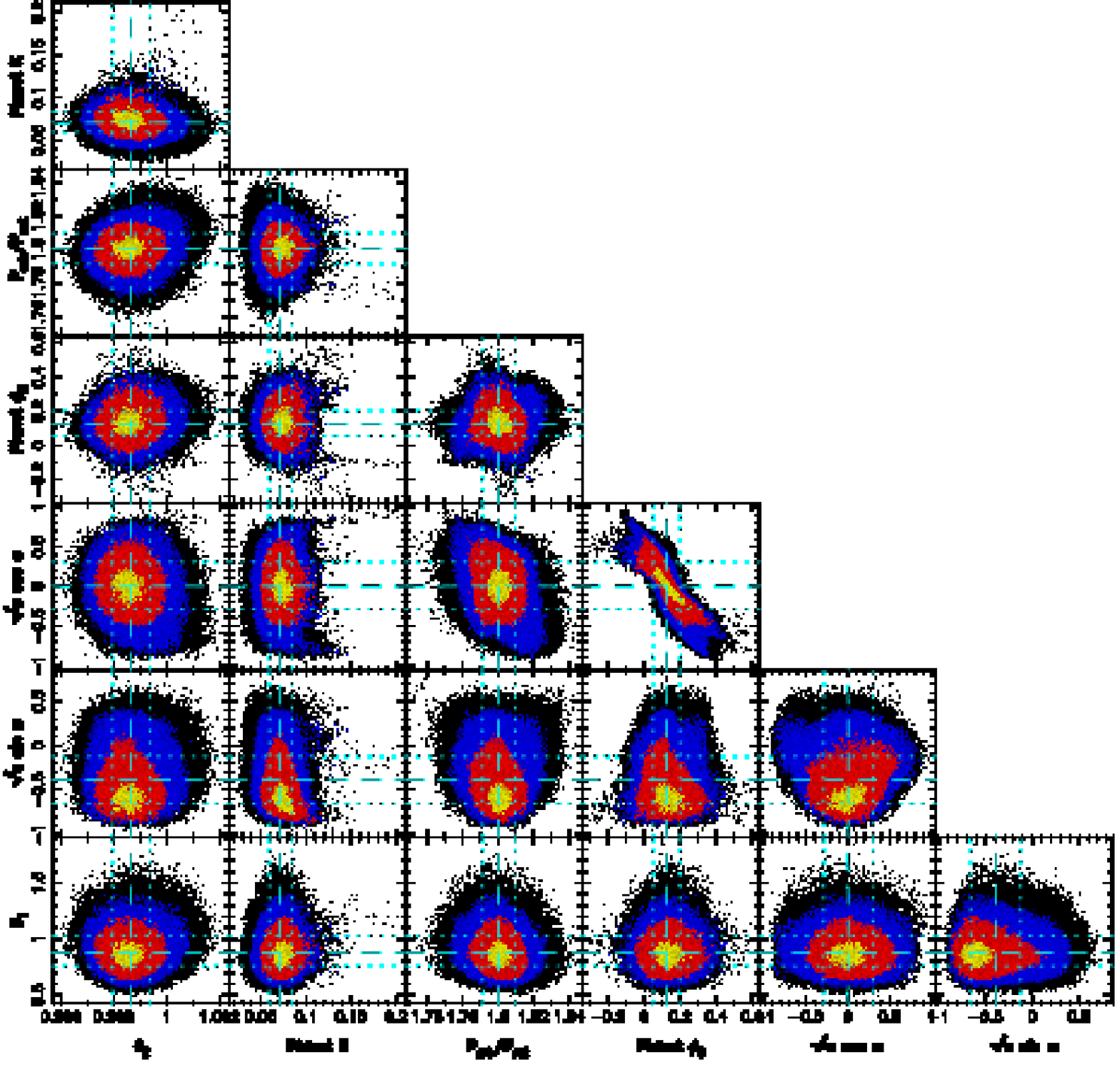}\hspace{3mm}
\includegraphics[scale=0.51,angle=-90]{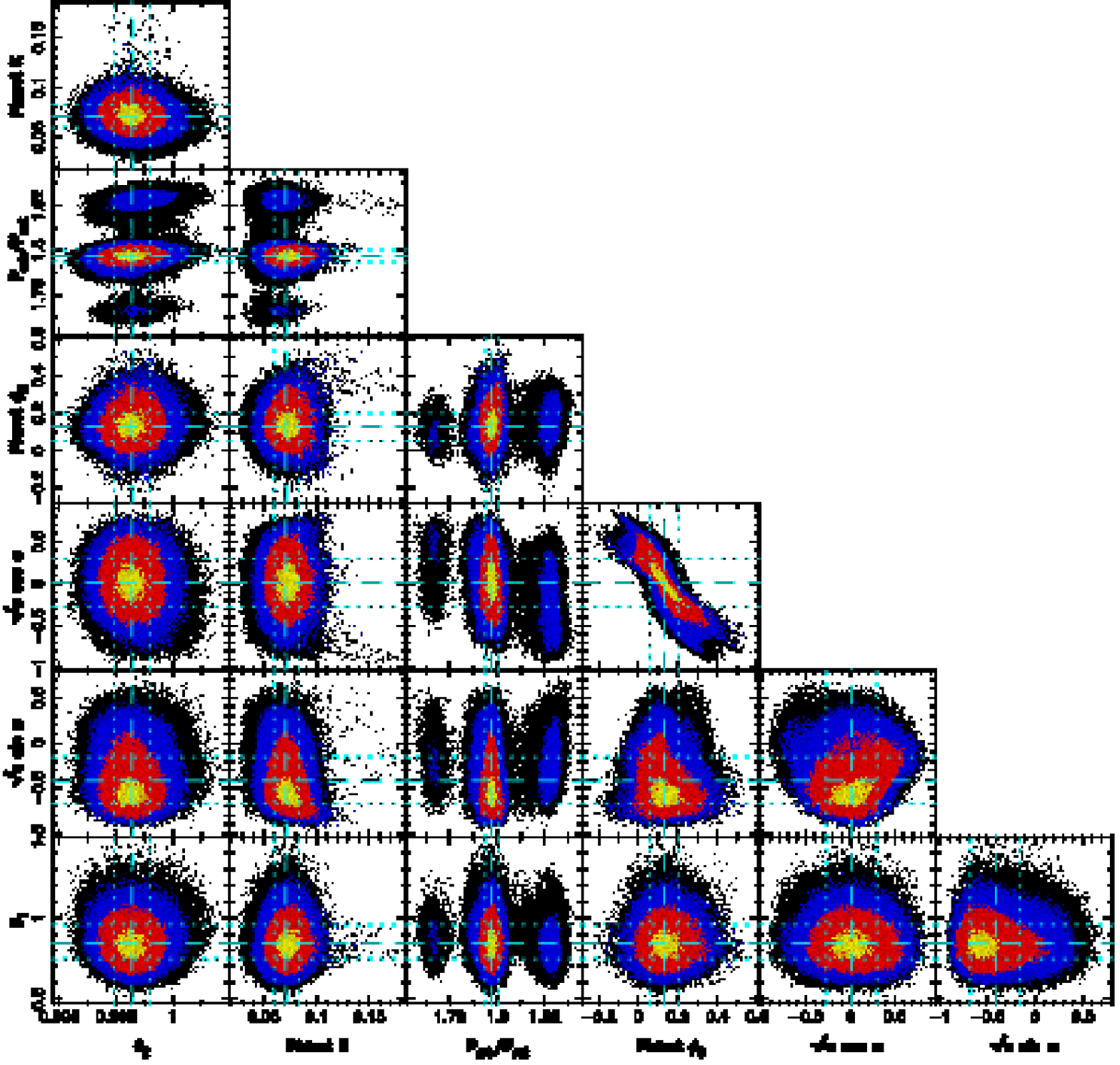}} 
\caption[]{Same as Fig.~\ref{fig:mcmc} when fitting an eccentric orbit using parameter $\sqrt{e} \cos \omega$ 
and $\sqrt{e} \sin \omega$ as search parameters where $e$ and $\omega$ respectively denote the eccentricity and argument of periapsis of the orbit;  
the marginal likelihood of the best eccentric orbit model is not significantly larger than that of the best circular orbit model, implying that 
there is no evidence that the planet is eccentric. } 
\label{fig:mcmc2}
\end{figure*}

\bibliography{v830tau}
\bibliographystyle{mnras}

\vspace{1cm}\small
\noindent $^1${\it Universit\'e de Toulouse, UPS-OMP, IRAP, 14 avenue E.~Belin, Toulouse, F--31400 France} \\
$^2${\it CNRS, IRAP / UMR 5277, Toulouse, 14 avenue E.~Belin, F--31400 France} \\
$^3${\it CFHT Corporation, 65-1238 Mamalahoa Hwy, Kamuela, Hawaii 96743, USA} \\
$^4${\it SUPA, School of Physics and Astronomy, Univ.\ of St~Andrews, St~Andrews, Scotland KY16 9SS, UK} \\
$^5${\it D\'epartement de physique, Universit\'e de Montr\'eal, C.P.~6128, Succursale Centre-Ville, Montr\'eal, QC, Canada  H3C 3J7} \\
$^6${\it Crimean Astrophysical Observatory, Nauchny, Crimea 298409} \\
$^7${\it Department of Physics and Astronomy, York University, Toronto, Ontario L3T 3R1, Canada} \\ 
$^8${\it ESO, Karl-Schwarzschild-Str.\ 2, D-85748 Garching, Germany} \\
$^9${\it School of Physics, Trinity College Dublin, the University of Dublin, Ireland} \\ 
$^{10}${\it Departamento de F\`{\i}sica -- ICEx -- UFMG, Av. Ant\^onio Carlos, 6627, 30270-901 Belo Horizonte, MG, Brazil} \\
$^{11}${\it Harvard-Smithsonian Center for Astrophysics, 60 Garden Street, Cambridge, MA 02138, USA} \\ 
$^{12}${\it Universit\'e Grenoble Alpes, IPAG, BP~53, F--38041 Grenoble C\'edex 09, France} \\
$^{13}${\it CNRS, IPAG / UMR 5274,  BP~53, F--38041 Grenoble C\'edex 09, France} \\
$^{14}${\it Institute of Astronomy and Astrophysics, Academia Sinica, PO Box 23-141, 106, Taipei, Taiwan} \\
$^{15}${\it Kavli Institute for Astronomy and Astrophysics, Peking University, Yi He Yuan Lu 5, Haidian Qu, Beijing 100871, China} \\
$^{16}${\it LUPM, Universit\'e de Montpellier, CNRS, place E.~Bataillon, F--34095 Montpellier, France} 

\bsp	
\label{lastpage}
\end{document}